\documentclass[fleqn,usenatbib]{mnras}

\usepackage{newtxtext,newtxmath}
\usepackage{lineno}

\usepackage[T1]{fontenc}

\DeclareRobustCommand{\VAN}[3]{#2}
\let\VANthebibliography\thebibliography
\def\thebibliography{\DeclareRobustCommand{\VAN}[3]{##3}\VANthebibliography}

\usepackage{graphicx}	
\usepackage{amsmath}	
\usepackage{caption}
\usepackage{subcaption}

\title[Survey of redback candidates with MeerKAT]{A targeted radio pulsar survey of redback candidates with MeerKAT}

\date{Accepted 2024 March 13. Received 2024 March 8; in original form 2023 December 15}

\pubyear{2024}

\begin{document}
\author[T. Thongmeearkom, et al. ]{T. Thongmeearkom$^{1,2}$\thanks{E-mail: tinn.thongmeearkom@postgrad.manchester.ac.uk}, C. J. Clark$^{3,4,1}$, R. P. Breton$^{1}$, M. Burgay$^{5}$, L. Nieder$^{3,4}$, P. C. C. Freire$^{6}$, \newauthor E.~D.~Barr$^{6}$, B. W. Stappers$^{1}$, S. M. Ransom$^{7}$, S. Buchner$^{8}$, F. Calore$^{9}$, D. J. Champion$^{6}$, \newauthor I. Cognard$^{10,11}$, J.-M. Grie{\ss}meier$^{10,11}$, M. Kramer$^{6,1}$, L. Levin$^{1}$, P. V. Padmanabh$^{3,4,6}$, \newauthor A. Possenti$^{5}$, A. Ridolfi$^{5,6}$, V. Venkatraman Krishnan$^{6}$ and  L. Vleeschower$^{1,12}$
\\
$^{1}$ Jodrell Bank Centre for Astrophysics, Department of Physics and Astronomy, The University of Manchester, Manchester M13 9PL, UK\\
$^{2}$ National Astronomical Research Institute of Thailand, Don Kaeo, Mae Rim, Chiang Mai 50180, Thailand\\
$^{3}$ Max Planck Institute for Gravitational Physics (Albert Einstein Institute), D-30167 Hannover, Germany\\
$^{4}$ Leibniz Universit\"{a}t Hannover, D-30167 Hannover, Germany\\
$^{5}$ INAF -- Osservatorio Astronomico di Cagliari, Via della Scienza 5, I-09047 Selargius (CA), Italy\\
$^{6}$ Max-Planck-Institut f\"{u}r Radioastronomie, Auf dem H\"{u}gel 69, D-53121 Bonn, Germany\\
$^{7}$ National Radio Astronomy Observatory, 520 Edgemont Rd., Charlottesville, VA 22903, USA\\
$^{8}$ South African Radio Astronomy Observatory, 2 Fir Street, Black River Park, Observatory 7925, South Africa\\
$^{9}$ LAPTh, CNRS, USMB, F-74940 Annecy, France\\
$^{10}$LPC2E - Universit\'{e} d'Orl\'{e}ans /  CNRS, 45071 Orl\'{e}ans cedex 2, France\\ 
$^{11}$Observatoire Radioastronomique de Nan\c{c}ay (ORN), Observatoire 
de Paris, Universit\'{e} PSL, Univ Orl\'{e}ans, CNRS, 18330 Nan\c{c}ay, France\\
$^{12}$Center for Gravitation, Cosmology, and Astrophysics, Department of Physics, University of Wisconsin-Milwaukee, P.O. Box 413, Milwaukee, WI 53201, USA\\}
\label{firstpage}
\pagerange{\pageref{firstpage}--\pageref{lastpage}}
\maketitle


\begin{abstract}
Redbacks are millisecond pulsar binaries with low mass, irradiated companions. These systems have a rich phenomenology that can be used to probe binary evolution models, pulsar wind physics, and the neutron star mass distribution. A number of high-confidence redback candidates have been identified through searches for variable optical and X-ray sources within the localisation regions of unidentified but pulsar-like \textit{Fermi}-LAT gamma-ray sources. However, these candidates remain unconfirmed until pulsations are detected. As part of the TRAPUM project, we searched for radio pulsations from six of these redback candidates with MeerKAT. We discovered three new radio millisecond pulsars, PSRs~J0838$-$2527, J0955$-$3947 and J2333$-$5526, confirming their redback nature. PSR~J0838$-$2827 remained undetected for two years after our discovery despite repeated observations, likely due to evaporated material absorbing the radio emission for long periods of time. While, to our knowledge, this system has not undergone a transition to an accreting state, the disappearance, likely caused by extreme eclipses, illustrates the transient nature of spider pulsars and the heavy selection bias in uncovering their radio population. Radio timing enabled the detection of gamma-ray pulsations from all three pulsars, from which we obtained 15-year timing solutions. All of these sources exhibit complex orbital period variations consistent with gravitational quadrupole moment variations in the companion stars. These timing solutions also constrain the binary mass ratios, allowing us to narrow down the pulsar masses. We find that PSR~J2333$-$5526 may have a neutron star mass in excess of 2\,M$_{\odot}$. 
\end{abstract}

\begin{keywords}
pulsars: general -- pulsars: individual: PSR~J0838$-$2827, PSR~J0955$-$3947, PSR~J2333$-$5526 -- binaries: general -- gamma rays: stars
\end{keywords}


\section{Introduction}
\label{S:intro}

Spider pulsars are a family of binary systems in which a millisecond pulsar (MSP) orbits with a tidally-locked, low-mass degenerate or semi-degenerate companion in a tight orbit \citep[$\sim$few hours up to a day; e.g.][]{Roberts_2017}. Their population is generally subdivided into two main categories, black widows and redbacks, according to the mass of their companion \citep[$\ll 0.1 {\rm M}_\odot$ and $0.2 - 0.4 {\rm M}_\odot$, respectively;][]{Roberts_2013}. In these systems, the two stars interact via an intra-binary shock between the pulsar’s relativistic wind and the companion stellar wind. The very energetic pulsar is responsible for the strong irradiation of the companion whereby the side that faces the pulsar can be heated to several thousands of degrees above the opposite side, thus creating strong variability at visible wavelengths \citep[e.g.][]{Callanan1995+B1957,Breton_2013}. Furthermore, in a large fraction of systems, material blown away from the companion's surface by the pulsar wind can obstruct the radio emission from the pulsar (most likely via synchrotron absorption) and cause eclipses that can extend for a significant fraction of the orbit \citep[e.g.][]{Polzin_2018,Polzin_2019}. Spider companions therefore `evaporate' over time, though current mass-loss rate estimates \citep[e.g.][]{Polzin_2018} suggest that this process is not sufficient to provide a formation channel for isolated MSPs, as initially proposed by \citet{Fruchter_1988}.

The evolutionary history of spider pulsars leading to their current state is still poorly understood, but they most certainly experience a phase of mass transfer from the companion, which pumps angular momentum into the pulsar and where the system would be visible as an X-ray binary \citep{Alpar_1982,Radhakrishnan_1982,Bhattacharya_1991}. This evolutionary link was clearly evidenced with the discovery of the accreting MSPs \citep{Wijnands_1998,Chakrabarty_2003} and then of the transitional millisecond pulsars in which X-ray binaries have been seen to turn into redback pulsars \citep{Archibald_2009} and vice versa \citep{Papitto_2013,Stappers_2014}. More recently, it became clearer that irradiation effects play a crucial role in dictating their evolution pathway, with the irradiation efficiency potentially determining if a system turns into a black widow or a redback \citep[e.g.][]{Chen_2013,DeVito_2020}.

Multi-wavelength studies make it possible to constrain orbital parameters and masses of spider pulsar systems. Most notably, pulsar timing in the radio and gamma-ray regimes enables one to measure the five Keplerian parameters of the pulsar with great accuracy. Optical photometric light curves, on the other hand, can be modelled to derive the orbital inclination and, if combined with radial velocity information from spectroscopy of the companion, they can be used to infer the mass of each component \citep{PSRHandbook}. Evidence showing that spider systems have higher-than-average neutron star masses \citep[$ 1.46\pm 0.30\,{\rm M}_\odot$ for an average neutron star;][]{Zhang2011+neutronstarmass} has now been found, with a report that the neutron stars in redbacks have a median mass of $1.78 \pm 0.09 {\rm M}_\odot$ \citep{Strader_2019}. This motivates searches for $> 2 {\rm M}_\odot$ neutron stars, which could constrain the equation of state. -- some interesting examples above $2 {\rm M}_\odot$ have possibly been found \citep{Linares_2019, Romani2022+J0952}.

Multiple approaches have been pursued to find new spider pulsars. Unguided widefield radio pulsar survey has only found a handful of the $\sim80$ known spiders thus far \citep[the ATNF Pulsar Catalogue;][]{psrcat}, including the original black widow \citep{Fruchter_1988}. This technique is comparatively less efficient at finding spider pulsars than other types of binary pulsars, primarily since radio eclipses and the relatively large and rapidly changing accelerations heavily hinder the detection of spiders. Instead, the bulk of the spider population has been uncovered through the targeted approach of pointing at candidates identified via multi-wavelength campaigns. By far the most successful approach to finding new Galactic spiders has been to pursue their gamma-ray signature. In 2008, the \textit{Fermi Gamma-ray Space Telescope} started to operate. One of the telescope's instruments, the Large Area Telescope \citep[LAT;][]{Atwood_2009}, is especially suitable for detecting the pulsed GeV gamma-ray emission from millisecond pulsars \citep{Abdo2009+MSPs}, which are particularly efficient high-energy emitters \citep{Smith2023+3PC}. The \textit{Fermi}-LAT has detected and localised a large number of point sources. The latest catalogue, the \textit{Fermi}-LAT Fourth Source Catalogue \citep{Abdollahi_2020} Data Release 4 (4FGL-DR4, \citealt{4FGLDR4}) detecting more than 7000 sources, while the Third \textit{Fermi}-LAT Gamma-ray Pulsar Catalogue contains nearly 300 confirmed gamma-ray pulsars \citep{Smith2023+3PC}. 

 Point-like gamma-ray sources that have not been associated with other objects at other wavelengths -- of which thousands remain -- provide a very effective list of locations that can be searched for energetic pulsars in the radio where they are bright enough to yield a detection over a short minutes-to-hour observation. Gamma-ray point source characteristics such as variability and spectral properties can also be used to identify sources most likely to be a pulsar \citep[e.g.][]{SazParkinson2016+ML}. Multiple surveys, initially coordinated through the \textit{Fermi} Pulsar Search Consortium \citep[PSC;][]{Ray2012+PSC}, have successfully employed this strategy, including studies conducted with the Green Bank Telescope (GBT) \citep{Ransom_2011}, the Murriyang Parkes telescope \citep[e.g.][]{Keith_2011,Camilo_2015}, the Arecibo telescope \citep{Cromartie_2016} and the Five-hundred-meter Aperture Spherical radio Telescope (FAST) \citep{Wang2021+FASTMSP}. Given that an important limitation to finding pulsars is instantaneous sensitivity, the MeerKAT radio telescope has recently provided us with a new, improved tool for this quest, as was demonstrated by the first results of the TRansients and PUlsars with MeerKAT (TRAPUM) survey of unidentified \textit{Fermi}-LAT sources, which discovered 9 new pulsars from a search of 79 unidentified gamma-ray sources \citep{Clark_2023}.

A similar strategy to the one laid out above is to pursue candidate spider pulsars identified through searches for periodic optical and X-ray counterparts, often confirmed with spectral velocity measurements indicating an unseen neutron-star-mass primary \citep[e.g.][]{Kong2012+J2339,Romani2012+J1311}. These candidates are usually identified through the monitoring of suitable \textit{Fermi} gamma-ray unidentified point sources or serendipitous observations of such fields made by large-scale surveys. More than a dozen candidate redback and black-widow binaries have been identified in this way (see Table 15 of \citealp{Smith2023+3PC}), but only a handful of these have been confirmed by a pulsation detection. Not including the discoveries that we present here, just three optically-identified redback candidates, PSRs~J2339$-$0533 \citep{Ray2020+J2339}, J0212$+$5320 \citep{Perez_2023} and J1910$-$5320 \citep{Au_2023+J1910,Dodge2024+1910} were confirmed via radio pulsations, while a further three binaries were confirmed by direct searches for gamma-ray pulsations in the \textit{Fermi}-LAT data \citep{Pletsch2012+J1311, Nieder2020+J1653, Clark2021+J2039}, using constraints from optical monitoring to reduce the orbital parameter space. The long integration time required to accumulate enough gamma-ray photons makes this latter method a challenging computational task \citep{Nieder2020+Methods}, and in many cases the available orbital constraints are not sufficiently precise to make a gamma-ray pulsation search feasible, leaving radio searches as the only means to confirm their MSP nature. However, only two of these three gamma-ray discovered spiders were eventually detected by folding radio observations using the gamma-ray timing ephemeris \citep{Ray2013+J1311,Corongiu2021+J2039}, and one of these, PSR~J1311$-$3430, is only very sporadically detectable. The third of these systems, PSR~J1653$-$0158, remains undetected in radio despite efforts by many large radio telescopes \citep{Nieder2020+J1653}, suggesting that some spider binaries may be nearly always enshrouded by evaporated material. Hence, both radio and gamma-ray searches of these candidate systems remain important complementary approaches for obtaining a more complete view of this population.

Despite the rather low yield to date, this refined strategy of targeting optically-identified candidate binaries provides a number of potential advantages over more traditional surveys. First, the optical signature provides a much stronger prior as to the existence of a spider pulsar than the gamma-ray properties. Second, the localisation of the candidate is generally much better (sub-arcsecond vs. arcminute uncertainties), thus allowing one to position the radio beam directly at the right location to avoid sensitivity losses that arise towards the edge of the beam. This is particularly crucial for MeerKAT observations, where hundreds of coherent tied-array beams with widths of a few arcsecs are required to tile the error box of an unassociated \textit{Fermi} source (which typically have widths measured in arcminutes), with typical sensitivity losses of around 50\% occurring midway between beams. The ensuing search for radio pulsations can also be operated on a smaller data set, which opens possibilities for longer pointings and searching at higher accelerations when processing resources are limited, as they are for regular TRAPUM observations. Finally, the optical observations also provide an approximate orbital period and reference phase for the system, which in turn can be used to schedule the radio searching to occur around the inferior conjunction of the pulsar when it is least likely to be eclipsed.

In this paper, we expand the quest to find more spider pulsars by focusing on six \textit{Fermi} unidentified gamma-ray sources, which have been found to have possible counterparts with redback-like optical light curves and/or modulated X-rays. Since these systems are likely to harbor a radio pulsar, we conducted a deep survey of these targets with the MeerKAT telescope, combining observations made at \textit{L}-band and Ultra High Frequency (UHF). The structure of the paper is as follows. Section \ref{S:survey} presents the survey strategy and targets. Section \ref{S:result} reports on the discoveries, gamma-ray searches and timing. Section \ref{S:discussion} discusses the characteristics of the discovered pulsars. Section \ref{S:conclusion} draws some conclusions and summarises the work.

\section{Survey and Observation}
\label{S:survey}
The MeerKAT telescope is an interferometer array that contains 64 radio dishes with a diameter of 13.5 metres. Located in the Karoo, a desert region in the Northern Cape province of South Africa, MeerKAT is the most sensitive telescope operating at centimetre wavelengths in the Southern Hemisphere. The location was chosen due to its low population density. Consequently, the site is only affected by very low levels of radio frequency interference (RFI) and is  high quality for radio astronomy \citep{Booth_2009}. At the time the survey was conducted, two receivers were available: an \textit{L}-band receiver operating between 856–1712 MHz, and a UHF receiver operating between 544–1088 MHz \citep[see][for more technical information]{Jonas_2016}.

Six targets have been observed as part of this survey (see Table \ref{T:6sources}). Multi-wavelength observations indicated they are very likely redback pulsar binaries. The \textit{Fermi}-LAT detected point sources in the direction of these sources that are not clearly associated to other known objects \citep{Abdollahi_2020}. Optical and X-ray searches of the few arc-minute localisation regions were conducted to shed light on their nature, and identified optical and/or X-ray counterparts with lightcurves displaying the typical orbital periodicity of redback systems \citep[e.g.][]{Halpern_2017,Li_2018,Swihart_2020}. The variability in the optical data can have either two maxima per orbital cycle, due to tidal distortion of the companion star, or one maximum per cycle if the pulsar's irradiation of the inner facing hemisphere of the companion star causes a large temperature difference. The X-ray modulation, on the other hand, is caused by an intra-binary shock and displays a double-peak pattern due to the emission from particles accelerated to high Lorentz factors along the shock boundary. Mass estimates derived from the projected radial velocity curve of the companions obtained from optical spectroscopy further indicate that these systems contain an unseen primary object with a mass compatible with that of a neutron star though no pulsations have been reported from these candidates until now \citep[e.g.][]{Rea_2017,Strader_2019}.

In light of the strong evidence provided by the combination of gamma-ray, X-ray and optical data, there is a very high likelihood that all six targets are redback binaries that harbour a millisecond radio pulsar. Some of these sources have been targeted several times in radio pulsation searches with other telescopes without success \citep[e.g.][]{Camilo_2015}. These non-detections could be due to the intrinsic faintness of the pulsars, but are may also be due to the eclipsing nature of these redback systems, many of which are only sporadically detectable \citep[e.g. PSR~J0212+5321 has been detected only twice despite years of searching; ][]{Perez_2023}. Therefore, we decided to pursue a deep radio search with MeerKAT in an attempt to detect radio pulsations from the pulsars that they may host.

\begin{table*}
	\caption{List of redback candidates in our survey, with parameter constraints from optical observations from the references. Positions are from \textit{Gaia} Data Release 3 \citep{Gaia,Gaia+DR3}. $T_{\rm asc}$ is the epoch of the ascending node of the pulsar in the Barycentric Dynamical Time (TDB) system. $P_{\rm b}$ is the orbital period. 1-$\sigma$ uncertainties on the final digits of the orbital parameters are quoted in parentheses. The $\textrm{DM}$ columns represent the predicted dispersion measures along these lines-of-sight at the distances estimated from previous studies, according to the NE2001 \citep{ne2001} and YMW16 \citep{YMW_2016} Galactic electron density models. The dagger sign ($\dagger$) represents the maximum predicted DM along the lines-of-sight from both models. $\textrm{DM}_{\rm search}$ are the DM ranges that we used for searching.}
	\label{T:6sources}
    \centering
    \renewcommand{\arraystretch}{1.0}
    \setlength\tabcolsep{3.0pt}
	\begin{tabular}{lccccccccc}
		\hline
		4FGL name& R.A. & Decl. & $T_{\rm asc}$ & $P_{\rm b}$  & d & $\textrm{DM}_{\rm NE2001}$ & $\textrm{DM}_{\rm YMW16}$  & $\textrm{DM}_{\rm search}$ & References\\
  & (J2000)& (J2000)& (MJD) & (d) & (kpc) & (pc cm$^{-3}$) & (pc cm$^{-3}$) & (pc cm$^{-3}$) & \\
		\hline
		J0523.3$-$2527 & $05^h23^m16\fs925$ & $-25\degr27\arcmin36\farcs92$ & 56577.3184(37) & 0.688135(3) & 2.24 & 40, 50$^{\dagger}$ & 33, 52$^{\dagger}$ & 100 & \citet{Strader_2014}\\
		J0838.7$-$2827 & $08^h38^m50\fs418$ & $-28\degr27\arcmin56\farcs97$ & 57781.2524(8) & 0.214523(5) & 1.70 & 117, 238$^{\dagger}$ & 242, 344$^{\dagger}$ & 250 & \citet{Halpern_2017}\\
		J0940.3$-$7610 & $09^h40^m23\fs787$ & $-76\degr10\arcmin00\farcs13$ & 58525.37263(172) & 0.270639(7) & 2.00 & 63, 110$^{\dagger}$ & 85, 137$^{\dagger}$ & 250 & \citet{Swihart_2021}\\
        J0955.3$-$3949 & $09^h55^m27\fs809$ & $-39\degr47\arcmin52\farcs30$ & 58097.44433(53) & 0.387339(8) & 1.70 & 88, 184$^{\dagger}$ & 217, 280$^{\dagger}$ & 350 & \citet{Li_2018}\\
        J1120.0$-$2204 & $11^h19^m58\fs309$ & $-22\degr04\arcmin56\farcs33$ & 59306.616(31) & 0.630398(27) & 1.00 & 24, 50$^{\dagger}$ & 28, 50$^{\dagger}$ & 60 & \citet{Swihart_2022}\\
        J2333.1$-$5527 & $23^h33^m15\fs968$ & $-55\degr26\arcmin21\farcs11$ & 58463.46881(65) & 0.287645(1) & 3.10 & 34, 35$^{\dagger}$ & 21, 23$^{\dagger}$ & 70 & \citet{Swihart_2020}\\
		\hline
	\end{tabular}
\end{table*}

\subsection{Targeted sources/ Redback Candidates}
\label{s:targetedsource}

\subsubsection{4FGL~J0523.3$-$2527}
\citet{Strader_2014} suggested that this LAT source is a new probable gamma-ray pulsar binary. This system has one of the largest inferred companion masses among known or candidate redbacks \citep{Strader_2019}, with a companion mass between 0.8$-$1.3 \(\textup{M}_\odot\) assuming a neutron star mass in the range 1.4$-$2.0 \(\textup{M}_\odot\). The companion has been classified as a late G or early K star using optical spectroscopy \citep{Strader_2014}. In addition to the projected radial velocity amplitude, rotational broadening was measured from the optical spectral lines, thus enabling a direct estimate of the mass ratio ($q = 0.61 \pm 0.06$) leading to the above companion mass estimate. The radial velocity measurement also shows evidence of eccentricity ($e = 0.04$) which would be the largest known in a redback system. Later, \citet{Halpern_2022} reported the most luminous optical and X-ray flares seen in a non-accreting pulsar so far. Previous attempts to find radio pulsations from this source have been unsuccessful \citep{Guillemot_2012,Petrov_2013}.

\subsubsection{4FGL~J0838.7$-$2827}
4FGL~J0838.7$-$2827 was identified as a high-confidence MSP candidate following an optical and X-ray spectral study, but with an unknown orbital period due to the limited duration of the observations \citep{Halpern_2017+0838MSP,Rea_2017}. The likely redback nature was confirmed later from improved photometry and spectroscopy, with the optical companion being a low-mass M dwarf star in a 5.15\,hr orbit around an unseen primary with a mass consistent with a neutron star \citep{Halpern_2017}. Flaring was detected similar to that seen in transitional MSPs, though the X-ray and gamma-ray luminosities suggest it is a non-accreting MSP binary \citep[e.g. $L_{\rm X-ray} \sim 10^{31} \textrm{erg s}^{-1}$ vs. $L_{\gamma} \sim 10^{33}\textrm{erg s}^{-1}$]{Halpern_2017+0838MSP}. The gamma-ray flux also shows no significant variability indicative of a transition over the 14-year \textit{Fermi}-LAT data set \citep{4FGLDR4}. The optical spectrum shows variable $\rm H\alpha$ emission, which is believed to come from the wind driven by the heated side of the companion as the emission line is too broad to be chromospheric in origin \citep{Halpern_2017}. Finally, The Australia Telescope Compact Array (ATCA) observations of the field put a $3\sigma$ upper limit of $\sim$20\,$\mu$Jy at 5.5 and 9\,GHz on the presence of a point source \citep{Rea_2017}.

\subsubsection{4FGL~J0940.3$-$7610}
4FGL~J0940.3$-$7610 was proposed as a redback candidate by \citet{Swihart_2021} based on the discovery of a variable optical and X-ray counterpart to the {\it Fermi}  point source. The optical light curve displays both ellipsoidal variations and irradiation with amplitudes typical of redback systems, with an orbital period of 6.5 hours. Orbital phase-resolved spectroscopy and modelling of the light curve suggest a lower mass neutron star ($1.2-1.4$ \(\textup{M}_\odot\)) and higher mass companion ($M_{\rm c} \gtrsim$ 0.4 \(\textup{M}_\odot\)) compared to what is typically seen in redbacks, however, without a pulsation detection or a reliable ephemeris to constrain the binary mass ratio, these values remain uncertain (see Section~\ref{S:mass} for further discussion). The model favours an edge-on orbit that suggests radio eclipses are very likely to occur in this system. \citet{Camilo_2015} performed a radio observation for this source with Murriyang but did not detect radio pulsations.

\subsubsection{4FGL~J0955.3$-$3949}
4FGL~J0955.3$-$3949 was noted as a strong MSP candidate based on a machine-learning classification of unidentified gamma-ray sources \citep{SazParkinson2016+ML}. A 9.3\,hr periodic optical variable was identified as a possible redback binary in the Catalina Surveys Southern periodic variable star catalogue \citep{Drake2017+CSSPVC}, with additional evidence found by \citet{Li_2018} via optical spectroscopy and X-ray variability. In addition, a 77~mJy pulsar-like radio counterpart was found by \citet{Frail_2016} in the 150\,MHz TGSS ADR catalogue \citep{Intema2017+TGSSADR}, but was discarded as it lies outside the 3FGL \textit{Fermi} localisation error \citep{3FGL} -- this source is, in fact, coincident with the optical counterpart, which lies well within the improved 4FGL localisation region \citep{Li_2018}. However, repeated radio follow-up observations by \citet{Camilo_2015} with Parkes, prior to the identification as a redback candidate, detected no radio pulsations, despite this source lying within the beam. 

\subsubsection{4FGL~J1120.0$-$2204}
4FGL~J1120.0$-$2204 was the second brightest unclassified source in the 4FGL DR3 catalogue \citep{4FGLDR3}. The presumed optical counterpart to the {\it Fermi} source, found by \citet{Swihart_2022}, features a binary companion with a mass around 0.17 \(\textup{M}_\odot\) in a sub-day orbital period ($P_{\rm b} \sim$15.1 hr) around an unseen neutron star. As suggested by these authors, this system might be at an intermediate stage leading the companion to form an exceptionally low-mass helium white dwarf orbiting a neutron star, e.g. akin to PSR~J1012+5307 \citep{mata2020}. Notably, optical observations have revealed a lack of variability in the system's behavior, suggesting the orbit might be seen with a face-on orientation. Furthermore, 4FGL~J1120.0$-$2204 displays a lower X-ray luminosity when compared to most redback systems. While it may not fit the conventional criteria for a redback candidate, its exceptional properties have led us to include 4FGL~J1120.0$-$2204 in our survey, recognising its importance for binary evolution and bridging the gap between spider and white dwarf pulsar systems.

No radio pulsar has been detected within this system despite extensive radio searches conducted with various telescopes, including Effelsberg \citep{Barr_2012}, GMRT \citep{Bhattacharyya2021+GMRT} and unpublished PSC searches with Parkes and GBT, while \citet{Hui2015+1120} inspected nearby radio sources from two catalogues, the Molonglo Sky Survey \citep{Bock_1999} and the NRAO VLA Sky Survey \citep{Condon_1998}. 

\subsubsection{4FGL~J2333.1$-$5527}
Our final candidate, 4FGL J2333.1$-$5527, exhibits an orbital period of approximately 6.9 hr and features a companion in a nearly edge-on orbit \citep{Swihart_2020}. The detection of a significant gamma-ray eclipse confirms this picture \citep{Clark2023_eclipses}. Optical photometry reveals a typical redback light curve, while optical spectroscopy indicates the presence of a mid-K type companion. Furthermore, the study suggests that the neutron star mass in this system likely exceeds 1.4 \(\textup{M}_\odot\), but it remains poorly constrained. No previous radio observations of 4FGL~J2333.1$-$5527 have been published. 

\subsection{Survey configuration}
We designed our campaign to observe each target for one hour as this integration time should provide a significant sensitivity improvement over previous efforts, yet a duration between $5-20$\% of the orbital periods means that pulsations can still be recovered from a search of the full time series, using only one or two derivatives to account for the orbital motion, i.e. the so-called acceleration \citep{JohnstonKulkarni+accel} and jerk searches \citep{Andersen_2018}, respectively. Each target was observed twice at both \textit{L}-band and UHF. These two bands provide complementary information; on the one hand, pulsars often display steep radio spectra which favour low-frequency observations \citep{Jankowski2018}, while on the other hand, radio eclipses tend to be shorter and less opaque at higher frequencies \citep[e.g.][]{Polzin_2020}. Similarly, repeated observations can help mitigate effects such as scintillation due to the scattering of the pulse in the interstellar medium \citep[e.g.][]{Camilo_2015}. Repeats avoid eclipses due to slightly different orbital phase coverage, and because eclipses, especially in their non-central part, vary greatly. Finally, our knowledge of the orbital periods from optical follow-up lets us plan observations when the pulsar is nearer inferior conjunction (i.e. between phases $0.5-1.0$ in the pulsar timing convention) where it is less likely to be eclipsed.

Note that 4FGL~J1120.0$-$2204 is an exception to the above strategy due its late addition to the list. We have only collected data from one epoch at each frequency at the time of writing this paper.

All 64 dishes were requested for our observations to maximise sensitivity. However, the number of dishes varied from one observation to another depending on the availability at that time (see Table \ref{T:epoch} and \ref{T:undetected_source}). Given that the optical counterparts provide accurate source positions, we could point a single tied-array beam at the candidates without worrying about beam size. We recorded the data using 4096 frequency channels and a time resolution of 38 $\mu$s and 60 $\mu$s at \textit{L}-band and UHF, respectively. This combination of frequency and time resolution maximises fast MSP detectability and mitigates the effect of the unknown dispersion measure (DM) along the line of sight, which can smear pulses within single channels, a particularly detrimental effect at short spin periods.

\subsection{Search tools and Acceleration search}
We employed three software packages: \texttt{PRESTO} \footnote{\url{https://github.com/scottransom/presto}}, \texttt{Pulsar\_miner} \footnote{\url{https://github.com/alex88ridolfi/PULSAR_MINER}} and \texttt{Spider\_twister} \footnote{\url{https://github.com/alex88ridolfi/SPIDER_TWISTER}}. \texttt{PRESTO} (PulsaR Exploration and Search TOolkit) is software for analysis and pulsar searching \citep{Ransom_2003}. \texttt{PULSAR\_MINER} is a python-based pipeline that automates all the steps of a PRESTO-based acceleration/jerk search (\texttt{DDplan} for DM steps, de-dispersion, searching, candidate filtering and folding), while \texttt{Spider\_twister} automatically searches in orbital phase with several trials of epoch of ascending node, $T_{\rm asc}$, to find the value that produces the highest signal-to-noise folded result using a preliminary ephemeris as a starting point.

In \texttt{PRESTO} the acceleration search range is parameterised by $z$, the (dimensionless) number of Fourier bins that a signal is smeared over due to orbital motion. This is related to the pulsar's acceleration, 
\begin{equation}
    a = \frac{zc}{hft_{\rm obs}^2} \, 
\label{E:accelsearch}
\end{equation}
where $f$ is the pulsar's fundamental spin frequency, $c$ is the speed of light, $T$ is the observation duration, $h$ is the harmonic number (such that the fundamental harmonic has $h=1$) and $a$ is the acceleration. A similar dimensionless parameter, $w$, quantifies the frequency smearing due to jerk, which is the first time derivative of the acceleration \citep{Andersen_2018}.
\begin{equation}
    \dot{a} = \frac{wc}{hft_{\rm obs}^3}
\end{equation}

Searching a range of non-zero jerks provides additional sensitivity to short period binaries (in the range $T \approx 0.1 P_{\rm b}$ to $T \approx 0.2 P_{\rm b}$), but requires more computational power than a pure acceleration search. In our study, we performed acceleration searches up to $z=200$ for all targets, and jerk searches up to $w=600$ for those with no clear detections in the acceleration search. Assuming a typical MSP spin period ($p=2$\,ms), these searches cover accelerations of 9.26\,m $\rm s^{-2}$, 37.03\,m $\rm s^{-2}$ and 333.33\,m $\rm s^{-2}$ for 1\,hr, 30\,min and 10\,min segments, respectively. We then calculated the maximum acceleration of 15 redbacks using the orbital parameters provided in \citet{Strader_2019}. Consequently, we found that the highest maximum acceleration from these 15 sources is 26\,m $\rm s^{-2}$ while the mean and the median are 11\,m $\rm s^{-2}$ and 8\,m $\rm s^{-2}$, respectively. We also limited the range of DMs to search by taking the larger of the maximum values predicted by the NE2001 \citep{ne2001} and YMW16 \citep{YMW_2016} models along the line of sight with extra ranges (see Table \ref{T:6sources}) for each target.

\section{Results}
\label{S:result}
\begin{figure*}
    \centering
	\includegraphics[width=19cm]{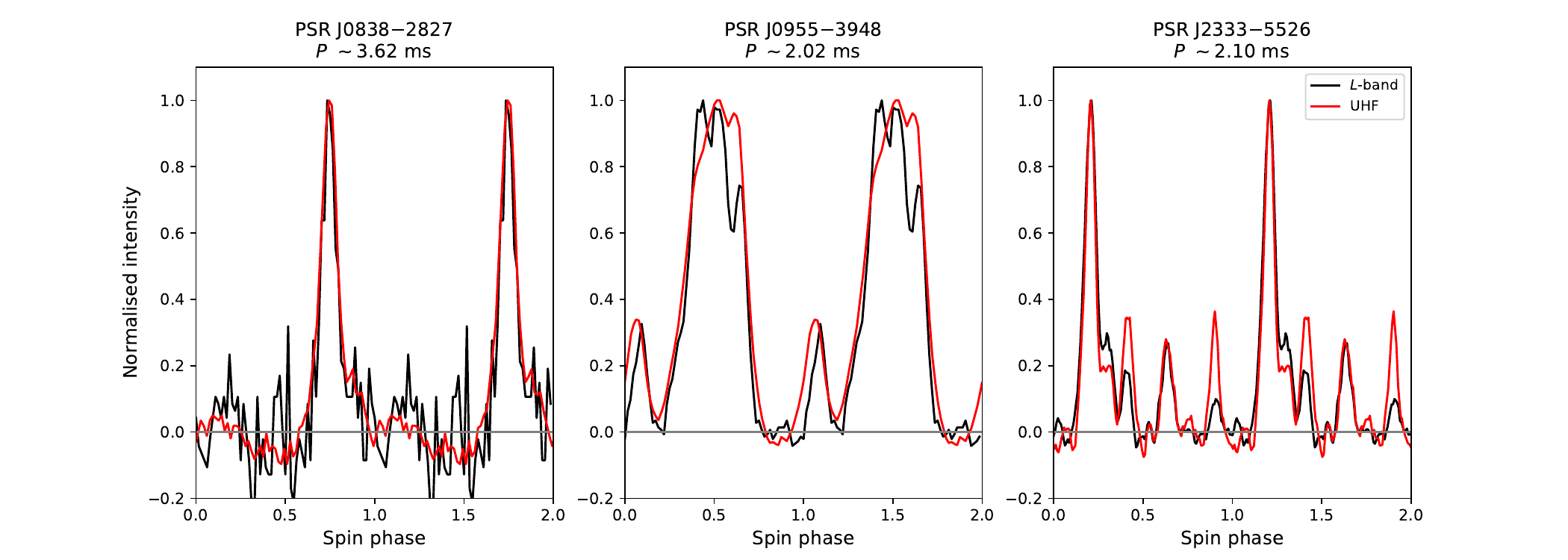}
    \caption{Pulse profiles of the three newly discovered redback pulsars at \textit{L}-band and UHF. The horizontal axis displays the spin phase of each source, replicated for clarity, while the vertical axis depicts the intensity, normalised to the peak value. The grey line presents the background level, calculated as the median of the off-pulse region. Since the gamma-ray ephemerides do not perfectly fold radio data to perform an absolute alignment, we aligned the \textit{L}-band (black) and the UHF (red) profiles using a cross-correlation.}
    \label{fig:3_detection}
\end{figure*}

Out of the six redback candidates we searched, we discovered a radio millisecond pulsar in three of them: PSRs~J0838$-$2827, J0955$-$3947 and J2333$-$5526 in 4FGL~J0838.7$-$2827, 4FGL~J0955.3$-$3949, and 4FGL~J2333.1$-$5527, respectively. All of them have been discovered at \textit{L}-band in 2021. Figure \ref{fig:3_detection} shows their pulse profiles at \textit{L}-band and at UHF. Accordingly, three new MSPs have spin periods of $P \sim$ 3.6, 2.0 and 2.1\,ms, with accelerations ($a\sim$13.6, 10.8 and 17.0\,m\,s$^{-2}$) demonstrating their binary nature. Further observations (see below) confirmed that their orbital periods are consistent with the optical counterparts. The spin periods are right in the range expected for redback systems, where most pulsars in these systems are found with $P \lesssim 5$\,ms \citep{Strader_2019}. 

\begin{figure}
	\includegraphics[width=6cm,height=14cm]{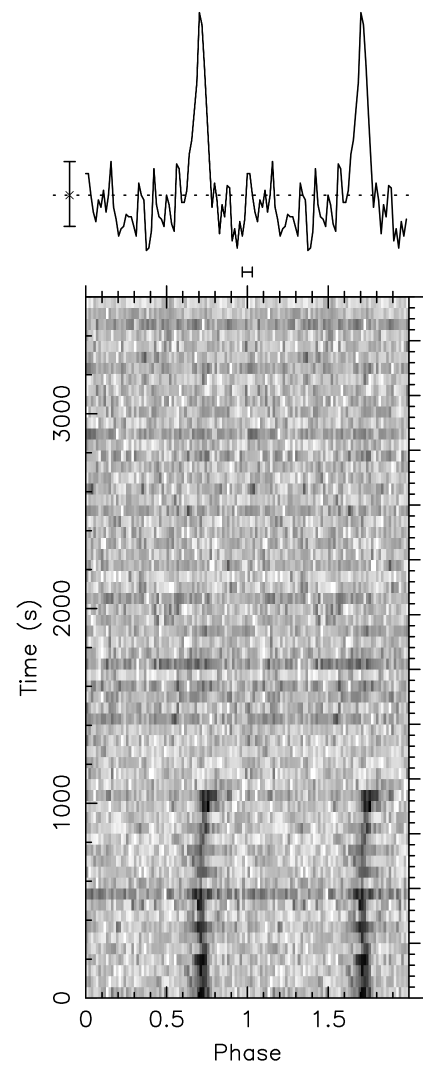}
    \centering
    \caption{Discovery observation of PSR~J0838$-$2827, with the folded pulse profile summed over frequency and time at the top, and the folded pulse profiles summed in frequency, but with sub-integrations running on the vertical axis at the bottom. The pulse cycle is replicated twice for clarity. In this \textit{L}-band observation dedispersed at $DM=80.55$ pc $\rm cm^{-3}$, the pulsations are clearly detected at the start before fading into a total eclipse around 1000\,s after the start. The pulse phase is also seen shifting due to the additional electron column density from the eclipsing medium increasing the DM temporarily.}
    \label{fig:J0838_eclipse}
\end{figure}

\begin{figure}
	\includegraphics[width=6cm,height=14cm]{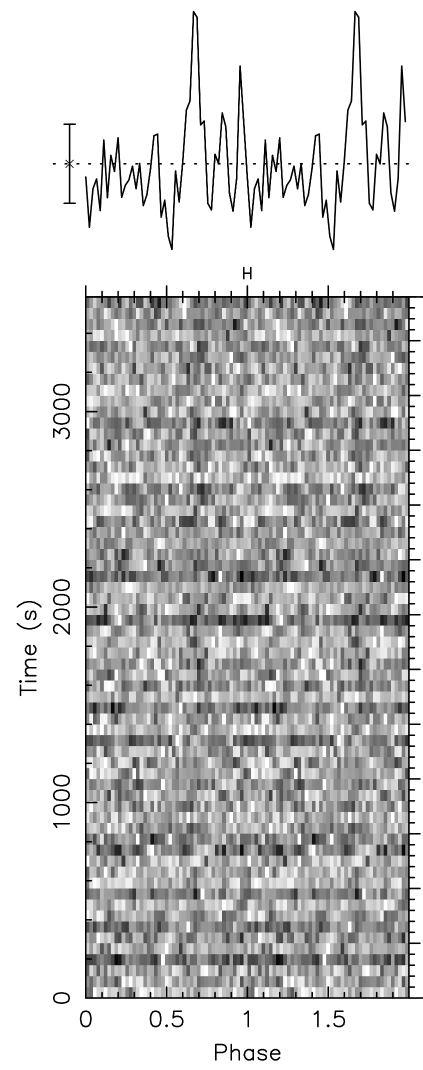}
    \centering
    \caption{Archival Parkes data of PSR~J0838$-$2827 folded using the timing parameters from MeerKAT (panels are as explained in Figure \ref{fig:J0838_eclipse}). A weak pulse can be seen later in the observation starting approximately 1500\,s into the observation after coming out of an eclipse.}
    \label{fig:old_parkes}
\end{figure}

\begin{figure}
    \centering
	\includegraphics[width=6cm,height=14cm]{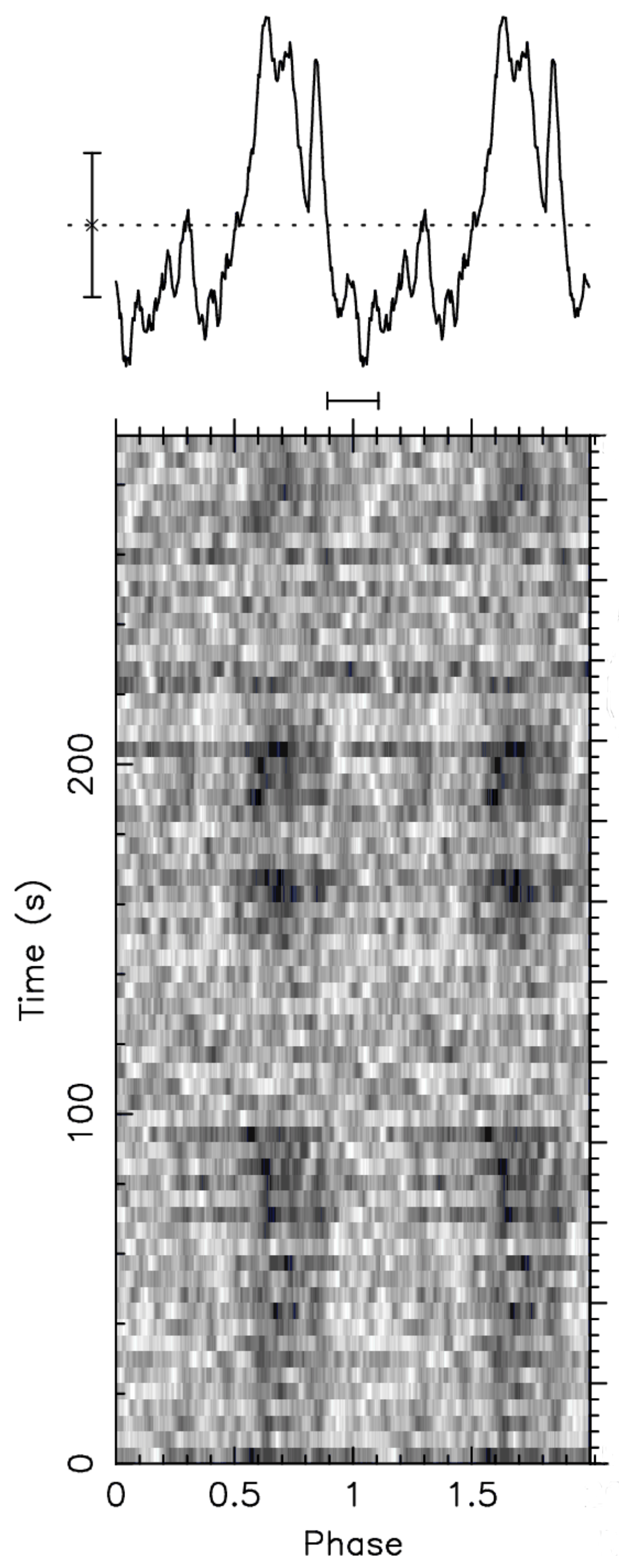}
    \caption{A five-minute MeerKAT \textit{L}-band timing observation of PSR~J0955$-$3947, beginning at orbital phase 0.05 (panels are as explained in Figure \ref{fig:J0838_eclipse}). Short eclipses can be seen throughout the observation around 100--140\,s and 220--260\,s.}
    \label{fig:J0955_mini_eclipse}
\end{figure}

Following these discoveries we searched archival observations from Murriyang Parkes and Green Bank telescopes that were accessible to us for the three new pulsars. Knowing the DM and the pulse period, and later having a partial timing solution, enabled us to detect all of them in a number of observations, with non-detections generally believed to result from eclipses or scattering (see Section~\ref{s:eclipses}). Full details of the epochs and phase coverage for the four 1\,hr observations and 5\,min timing campaign observations conducted with MeerKAT as well as ancillary data obtained from other facilities are provided in Table \ref{T:epoch}. Examples of the discovery, the archival data and the detection with eclipses are shown in Figure \ref{fig:J0838_eclipse}, \ref{fig:old_parkes} and \ref{fig:J0955_mini_eclipse}, respectively.

\subsection{Timing}
\label{s:Timing}
We undertook timing campaigns for each newly detected pulsar to obtain long-term phase-connected timing solutions that precisely describe their spin and orbital properties. These observations were performed at MeerKAT, using the Pulsar Timing User Supplied Equipment (PTUSE) backend \citep{Bailes2020+PTUSE}, with the 64\,m Murriyang Parkes telescope, using the Ultra Wide-Frequency Low receiver \citep[UWL;][]{Hobbs2020+UWL} and the Medusa Backend \citep[see Section 3.1 of][]{Hobbs2020+UWL}, and, for the northernmost and more elusive PSR~J0838$-$2827, also the Nan\c cay Radio telescope (NRT) located in France. 

Initially, we adopted a pseudo-logarithmic cadence strategy using the Parkes telescope (as time to follow up these sources at MeerKAT was granted at a later time), with one or two observations taken on the first day, followed by single observations on the second, fifth, tenth days, and so on. Typically, these observations lasted 1 to 2\,hr, depending on the target and, in case of a detection, provided one to four times-of-arrival (ToAs). 

While PSR~J0955$-$3949 was always detected, PSR~J2333$-$5526 was successfully detected more sporadically and PSR~J0838$-$2827 was never seen in our early UWL Parkes data, despite being visible in two archival observations obtained with a less sensitive backend. In October 2023, we then detected it again five times with Parkes.

A second pseudo-logarithmic campaign was hence carried out at MeerKAT on these two latter pulsars, with multiple 5\,min observations taken on day 1, followed by single observations in the following days as described above. These allowed us, in the case of PSR~J2333$-$5526, to get a good enough initial orbital solution, as described below, to properly fold previous Parkes data and, in some cases, recover the signal. PSR~J0838$-$2827 remained undetected in our MeerKAT timing campaign until one detection was made in each of August, September and October 2023. 

To follow up the sources over longer timescales, we continued our timing campaign with approximately monthly observations using the Parkes telescope and NRT.

We used the optical constraints given in Table~\ref{T:6sources} as starting points for solving the orbits and fitting a circular orbit model (using \texttt{PRESTO's} \texttt{fit\_circular\_orbit.py}) to the best-fitting spin frequency in each observation to obtain an initial estimate for the pulsar's orbital semi-major axis and to improve upon the optical estimates for $P_{\rm b}$ and $T_{\rm asc}$. After obtaining an acceptable initial orbit, we used \texttt{Dracula}\footnote{\url{https://github.com/pfreire163/Dracula}} \citep{Freire_2018} to determine the global rotation count of these pulsars and derive fully phase-connected timing solutions. Several orbital frequency derivatives needed to be added at this stage to obtain flat timing residuals with a reduced chi-square close to one and account for variations in the orbital period, a behaviour commonly seen in redback binaries \citep[e.g.][]{Pletsch2015+J2339,Deneva2016+J1048}.

\begin{figure*}
    \centering
	\includegraphics[width=18 cm]{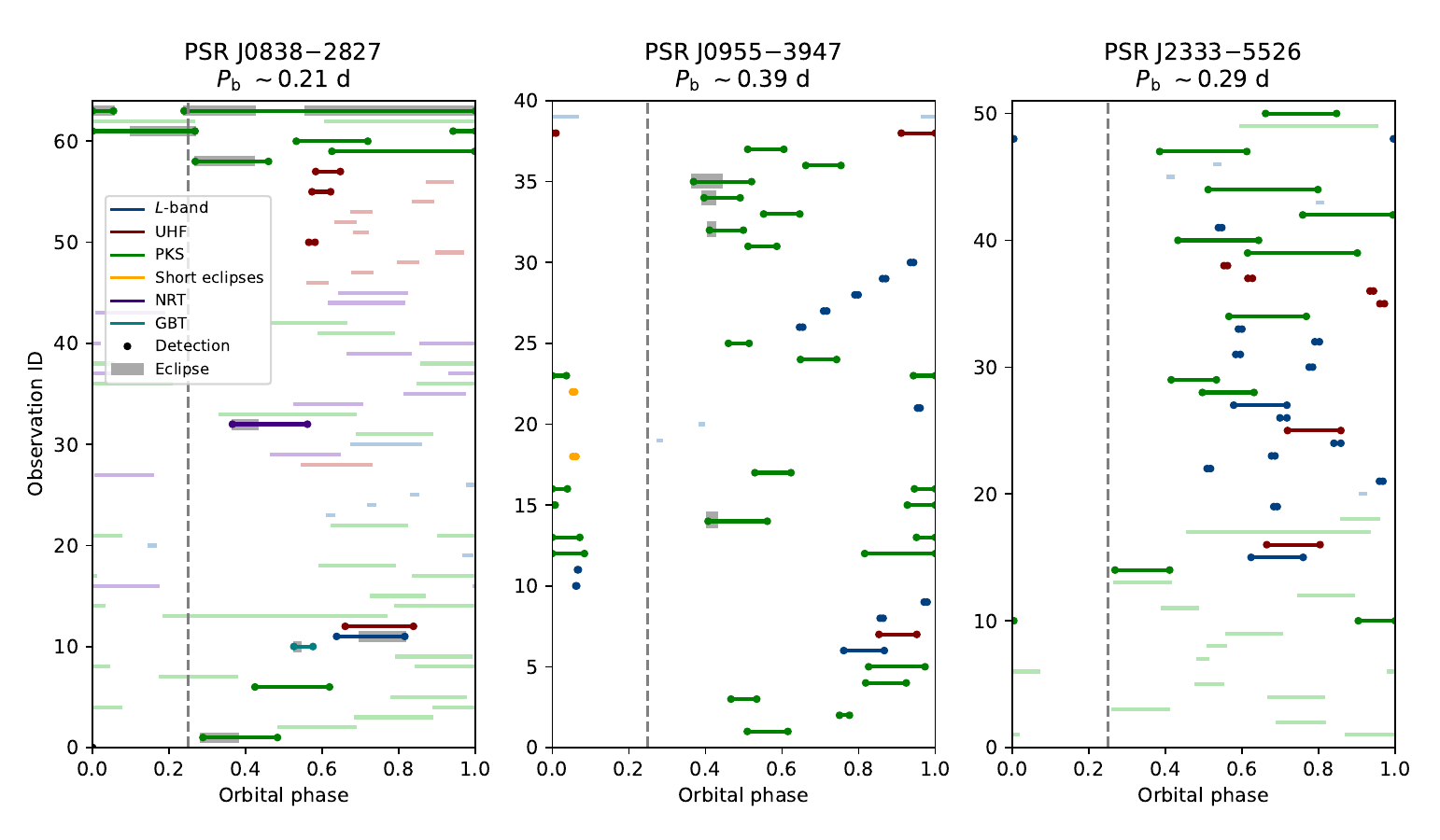}
    \caption{The orbital phase coverage of the MeerKAT \textit{L}-band, UHF, Parkes (PKS), GBT and NRT observation. Phase 0 corresponds to the pulsar at ascending node, and phase 0.25 is superior conjunction. The colours represent observations in the following bands/telescopes: UHF (red), L-band (blue), PKS (green), GBT (teal), NRT (purple), and \textit{L}-band with short eclipses observed (yellow). Detections are displayed by brighter colors with dots, while non-detection observations use dimmer colour without dots. The grey region is when the eclipse clearly shows up in the observation. Eclipses in \textit{L}-band and PKS archival data are shown in Figure \ref{fig:J0838_eclipse} and \ref{fig:old_parkes}. An example of short eclipses is in Figure \ref{fig:J0955_mini_eclipse}.}
    \label{fig:phase_cover}
\end{figure*}

Phase-connected radio timing solutions were obtained in this way for two of the three pulsars, PSRs~J0955$-$3947 and J2333$-$5526 (see Table~\ref{T:timing}). We used these as a basis to obtain a longer-term timing solution from the \textit{Fermi}-LAT gamma-ray data in Section~\ref{s:gamma-ray}. PSR~J0838$-$2827 has only been detected 14 times (see Section~\ref{s:eclipses} and Figure~\ref{fig:phase_cover}). Due to the long time separation of PSR~J0838$-$2827 between 2021 data and 2023 data, it was impossible to obtain a phase-connected timing solution covering all radio data. Therefore, we used only observations in 2023 for radio timing to create a starting point for gamma-ray timing (see Section \ref{s:gamma-ray}) for this source.

\begin{table*}
	\caption{Timing solutions for the newly discovered millisecond pulsars. Timing parameter values are from gamma-ray timing, described in Section~\ref{s:gamma-ray} using priors on astrometry parameters from \textit{Gaia}, and the radio timing value as a prior on $A_1$. \textbf{References:} $^a$ \citet{Halpern_2017}, $^b$ \citet{Li_2018}, $^c$ Strader, private communication (see Section \ref{S:mass}).}
	\label{T:timing}
	\centering
    \renewcommand{\arraystretch}{1.25}
    \begin{tabular}{lccc}
		\hline
		  Parameter & PSR~J0838$-$2827 & PSR~J0955$-$3947 & PSR~J2333$-$5526\\
		\hline
        \multicolumn{4}{c}{Timing parameters}\\
        \hline
		  R.A., $\alpha$ (J2000) & $08^h38^m50\fs41814(1)$ & $09^h55^m27\fs808690(6)$ & $23^h33^m15\fs96760(2)$\\
        Decl., $\delta$ (J2000) & $-28\degr27\arcmin56\farcs9729(3)$ & $-39\degr47\arcmin52\farcs2931(1)$ & $-55\degr26\arcmin21\farcs1056(4)$\\
        Proper motion in $\alpha$, $\mu_\alpha \cos \delta$ (mas yr$^{-1}$) & 1.5(3) & $-9.0(1)$& $0.6(3)$\\
        Proper motion in $\delta$, $\mu_\delta$ (mas yr$^{-1}$) & -12.3(3) & $6.1(1)$ & $-3.2(4)$\\
        Parallax, $\varpi$ (mas) & 0.43(40) & $0.27(13)$ & $-0.37(52)$\\
        Reference epoch for astrometry (MJD) & 57388 & 57388 & 57388 \\
        Dispersion Measure, DM (pc $\rm cm^{-3}$) & 80.656 & 130.711 & 19.925\\
        Spin frequency, $f$ (Hz) & 276.59755044807(9) & 494.38228117783(3)& 475.63324237057(1)\\
        Spin-down rate, $\Dot{f}$ (Hz $\rm s^{-1}$) & $-7.98(1)\times10^{-16}$ & $-9.3203(2)\times 10^{-15}$ & $-1.7647(2) \times 10^{-15}$\\
        2nd spin frequency derivative, $\ddot{f}$, (Hz $\rm s^{-2}$) & $5(4)\times10^{-26}$ & --- & ---\\
        3rd spin frequency derivative, $\dddot{f}$, (Hz $\rm s^{-3}$) & $-9(3)\times10^{-34}$ & --- & ---\\
        4th spin frequency derivative, $\ddddot{f}$, (Hz $\rm s^{-4}$) & $-4(1)\times10^{-41}$ & --- & ---\\
        Reference epoch for spin frequency (MJD) & 57461 & 59609 & 57500 \\
        Orbital period, $P_{\rm b}$ (d) & 0.214522813(5)  & 0.3873274(1) & 0.28764521(2)\\
        Projected semi-major axis, $A_1$ (lt s) & 0.43880(1) & 1.089707(4) & 0.896784(3)\\
        Epoch of ascending node, $T_{\rm asc}$ (MJD) & 60237.96601(7) & 58335.6527(4) & 57499.86418(6)\\
        \hline
        \multicolumn{4}{c}{Hyperparameters of orbital phase covariance function}\\
        \hline
		  Amplitude of orbital phase variations, $h$ (s) & $6^{+26}_{-2}$ &  $> 60$ & $16^{+19}_{-5}$ \\
        Length scale of orbital phase variations, $\ell$ (d) & $>200$ & $>2000$ & $450^{+360}_{-140}$ \\
        Mat\'{e}rn degree, $\nu$ & $>0.7$ & $1.40^{+0.24}_{-0.16}$ & $> 2.2$\\
        \hline
        \multicolumn{4}{c}{Derived parameters}\\
        \hline
        Spin period, $P$ (ms) & $3.615361012345(1)$ & 2.0227262142517(1) & 2.10246028014350(5)\\
        Spin period derivative, $\Dot{P}$ & $1.043(1)\times10^{-23}$ & $3.81333(7)\times10^{-20}$ & $7.8004(9)\times10^{-21}$\\
        Intrinsic spin-down rate, $\dot{f}_{\rm int}$ (Hz s$^{-1}$) & $-6.5\times10^{-16}$ & $-9.0\times10^{-15}$ & $-1.9\times10^{-15}$ \\
        Spin-down power, $\dot{E}$ (erg $\rm s^{-1}$) & $7.1\times10^{33}$ & $1.8\times10^{35}$  & $3.6\times10^{34}$\\
        Surface magnetic field strength, $B_{\rm S}$ (G) & $1.8\times10^{8}$ & $2.8\times10^8$ & $1.4 \times 10^8$\\
        Light-cylinder magnetic field strength, $B_{\rm LC}$ (G) & $3.5\times10^{4}$ & $3.1 \times 10^4$ & $1.3 \times 10^5$\\
        Nominal distance for Doppler corrections (kpc) & 1.7 & 3.5 & 3.1 \\
        Companion radial velocity, $K_{2} (\rm km \rm\,s^{-1})$ & 315(17)$^a$ & 272(4)$^b$ & 363(8)$^c$\\
        Inclination, $i (\degr)$ & $>78$ & $<76$ & $>78$\\
        Pulsar mass, $M_{\rm psr} ({\rm M}_\odot$) & 0.93(14) & 1.33(5) & 2.06(13)\\
        Companion mass, $M_{\rm c} ({\rm M}_\odot$) & 0.13(1) & 0.30(1) & 0.38(2)\\
        Mass ratio, $q = M_{\rm psr}/M_{\rm c}$ & 7.06(38) & 4.43(7) & 5.34(12)\\
		\hline
	\end{tabular}
\end{table*}

\subsection{Gamma-ray pulsation searches and timing}
\label{s:gamma-ray}
We used the radio timing solutions obtained in Section~\ref{s:Timing}, to search for gamma-ray pulsations from each new pulsar in the \textit{Fermi}-LAT data. We analysed \texttt{SOURCE}-class events using the ``Pass 8'' \texttt{P8R3\_SOURCE\_V3} instrument response functions \citep{Pass8,Bruel2018+P305}. We included photons according to the energy-dependent cuts that are described in Table 1 of \citet{4FGLDR3}. We used the \texttt{gtsrcprob} tool and the 4FGL-DR3 \citep{4FGLDR3} spectral and spatial models, with the \texttt{gll\_iem\_v07.fits} Galactic interstellar \citep{Fermi_IEM} and \texttt{iso\_P8R3\_SOURCE\_V3\_v1.txt} isotropic diffuse models, to assign weights representing the probability of each photon being emitted by the pulsar \citep{Kerr_2011}, as opposed to nearby point sources or the diffuse background.

Significant gamma-ray pulsations, with weighted H-test values \citep{Kerr_2011} of $H=71$ and $H=45$ were detected from PSRs~J0955$-$3947 and J2333$-$5526, respectively, within the validity period of the preliminary radio timing ephemerides. However, these timing solutions did not reveal pulsations in earlier \textit{Fermi}-LAT data. This is because the radio timing ephemerides for these pulsars begin at our discovery epochs, and contain several orbital frequency derivatives describing the orbital phase variations over time, and this Taylor series model extrapolates poorly to earlier times.  For PSR~J0838$-$2827, the initial radio ephemeris did not cover enough \textit{Fermi}-LAT data to reveal significant pulsations, but it did provide precise measurements of the pulsar's semi-major axis, and ascending node epoch. With these values fixed, and astrometry fixed at the \textit{Gaia} DR3 values for the companion star, we performed a search over the spin frequency, spin-down rate and orbital period. This led to a significant detection with $H=78$, but the pulse profile was unclear in some parts of the data, again indicative of orbital phase variations not accounted for in our initial timing model. 

\begin{figure*}
    \centering
    \includegraphics[width=1.0\textwidth]{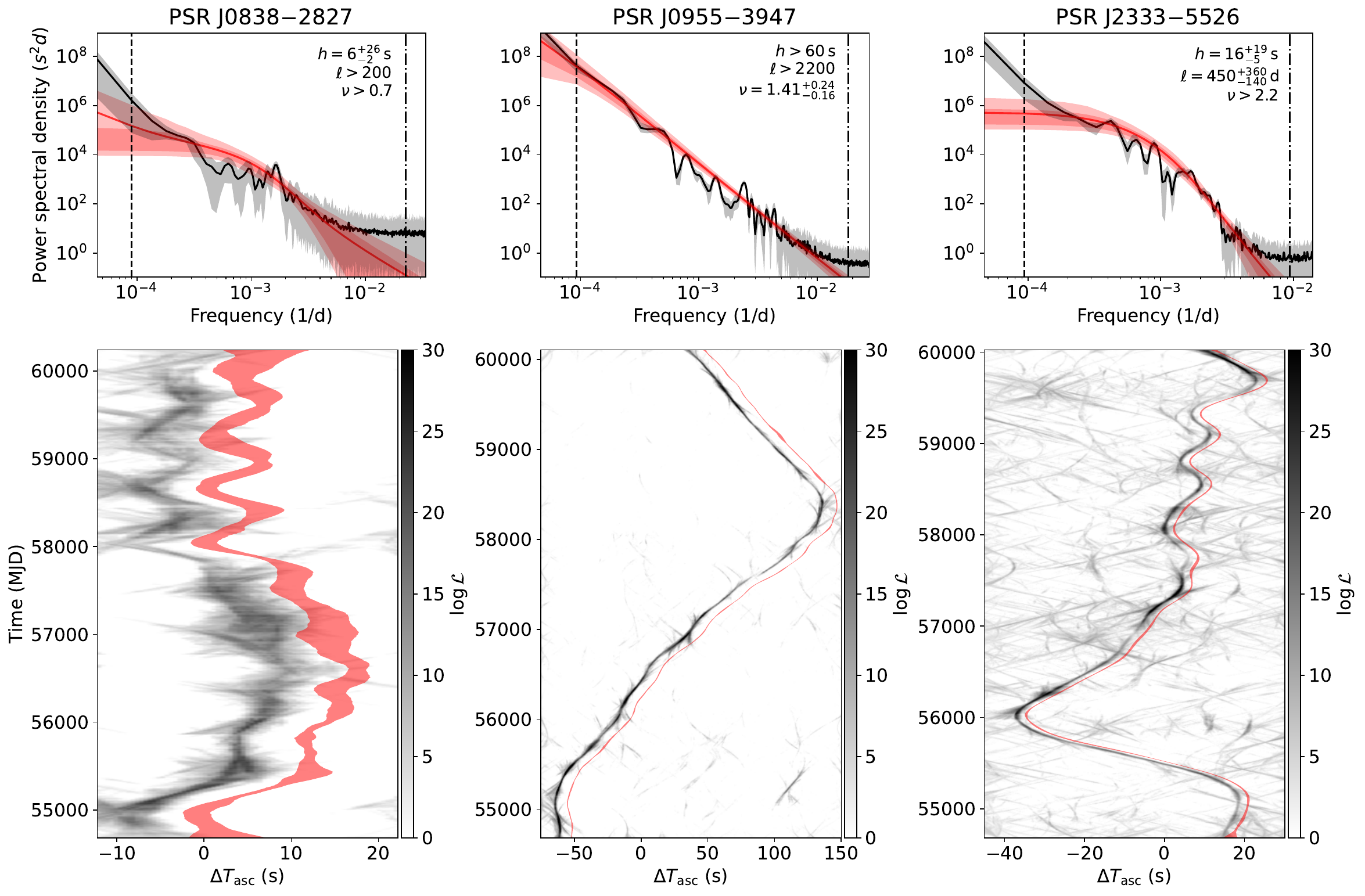}
    \caption{Orbital phase variations over the course of the \textit{Fermi}-LAT data.
    Upper panels show the power spectral densities (PSD) of the orbital phase variations. Red curves and shaded regions show the PSDs of the best-fitting Mat\'{e}rn covariance model, and 1 and 2-sigma uncertainties estimated from the Monte-Carlo samples, respectively. Black curves and grey shaded regions show the measured PSDs and their 95\% confidence intervals. The measured PSDs break to higher values than the Mat\'{e}rn models at high frequencies due to measurement uncertainties, and at low frequencies due to uncertainties caused by jointly fitting for the underlying orbital period and phase. Dashed and dot-dashed vertical lines show the minimum and maximum frequencies that were included in the timing model, respectively. Lower plots show the orbital phase variations as a function of time. The greyscale image shows the measured pulsation log-likelihood as a function of offsets from the pulsar's ascending node epoch, measured in overlapping 300-day intervals. Red curves show the 95\% confidence interval on the deviations in the pulsar's ascending node time, obtained from the Monte-Carlo samples, with a small offset for clarity. }
    \label{fig:OPV_plots}
\end{figure*}

To obtain a gamma-ray timing solution describing the entire 15-year \textit{Fermi}-LAT data set, we therefore first had to search over the orbital phase, period and the first orbital period derivative as a function of time, searching for coherent pulsations in overlapping 400-day intervals covering the LAT data. The results of these ``sliding window'' searches are shown in Figure~\ref{fig:OPV_plots}, revealing shifts in the pulsar's ascending node time of over 100\,s (for PSR~J0955$-$3947) compared to a model with a constant orbital period.

We then used the resulting orbital-phase vs. time measurements as a starting point to fit a fully-coherent timing model. We used the framework introduced in \citet{Clark2021+J2039} in which we modeled the orbital phase variations as a stochastic Gaussian process, and jointly fit the parameters of the covariance function of this process along with the standard timing model parameters, and those of the template pulse profile used to evaluate the likelihood of the photon-weighted pulse profile \citep{Kerr2015}. We use a Mat\'{e}rn covariance function, parameterised by an amplitude ($h$), length scale ($\ell$) and ``smoothness'' ($\nu$) hyperparameters. A process whose covariance function has these parameters has a smoothly broken power law power spectral density, which is flat below a cut-off frequency $\propto \ell^{-1}$, turning over to a power-law with spectral index $\gamma = -(2\nu + 1)$.  

\begin{figure*}
      \centering
	   \includegraphics[width=1.0\textwidth]{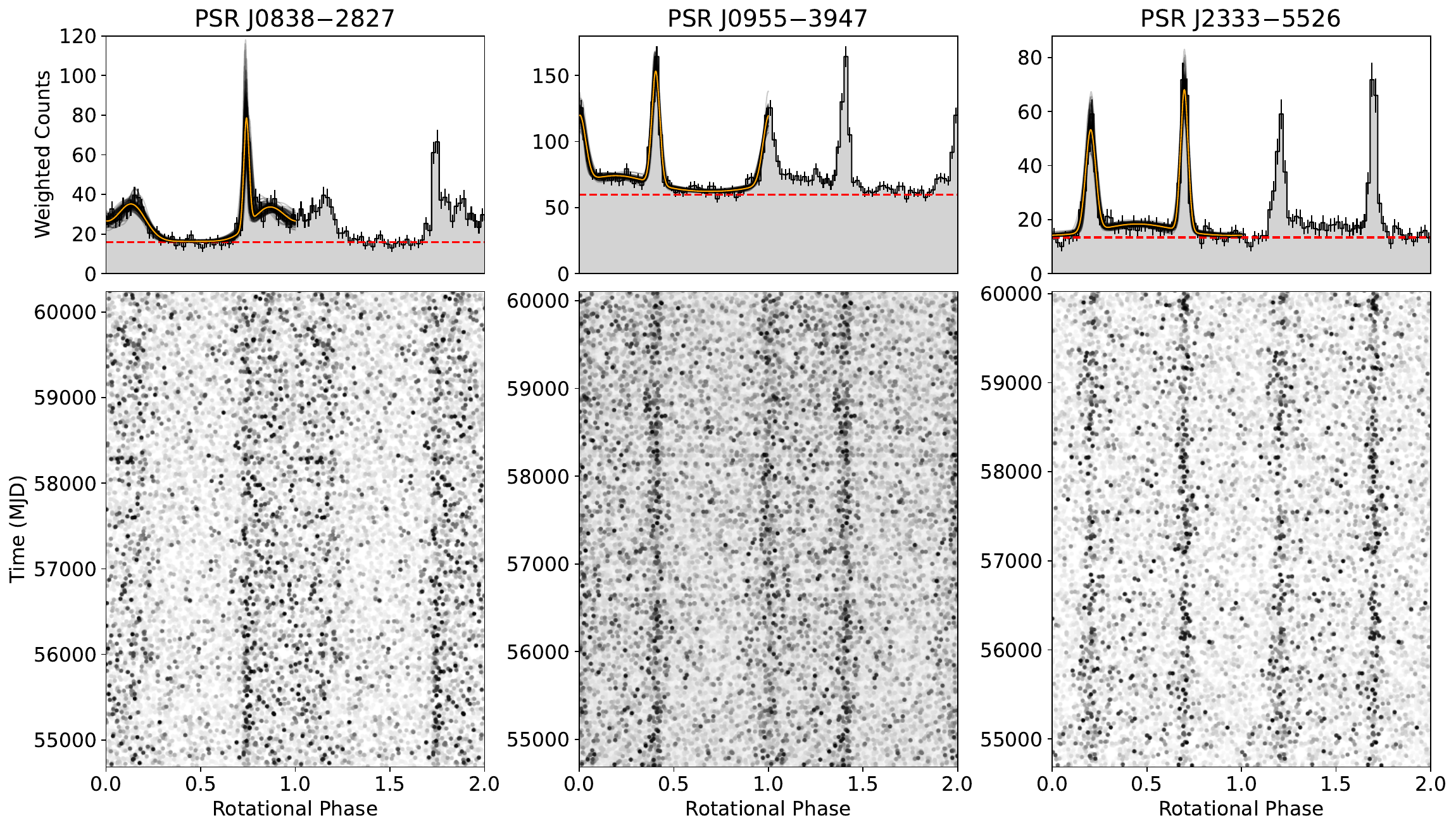}
    \caption{Gamma-ray pulsations from the three new redback MSPs. Lower panels show the weighted gamma-ray photon phases according to the best-fitting timing solution, while upper panels show the integrated pulse profile. The best-fitting template pulse profile is shown by an orange curve, with faint black curves underneath showing templates randomly drawn from the Monte-Carlo samples to illustrate the uncertainty on the template pulse profile.
    }
    \label{fig:photon_phases}
\end{figure*}

To make this procedure tractable, the method described in \citet{Clark2021+J2039} approximates the spin-phase likelihood function for each photon as a Gaussian, but we found that this approximation does not work well for fainter pulsars, or those with larger orbital phase variations. We therefore modified this procedure, using a Gibbs sampling method to deal with the multi-modal pulse profile. In this method, each photon is randomly assigned, according to its weight and an initial timing model, to either a single Gaussian pulse profile component (either one of two sharp peaks or to a bridge component connecting the two) or to the unpulsed background. The timing model (including the orbital phase variation Gaussian process) can then be fit analytically using generalised least squares, given these component assignments. The component assignments can then be updated according to a random sample drawn from the resulting posterior distribution on the timing model and template parameters, and this process is then repeated to generate a Monte-Carlo chain that approximates the full joint posterior on the timing model parameters, template pulse profiles, and Gaussian process hyperparameters. In the fitting, we adopted the \textit{Gaia} DR3 \citep{Gaia,Gaia+DR3} values and uncertainties as Gaussian priors on the astrometric parameters, and used radio timing estimates as priors on the pulsar semi-major axes. Gamma-ray timing is generally not precise enough to improve upon these measurements, with the exception of proper motion in PSR~J2333$-$5526 and the semi-major axis of PSR~J0838$-$2527, where the posterior uncertainties are slightly narrower. The estimated power spectral density of the orbital phase variations, and the best-fitting timing models from this procedure, are shown in Figures~\ref{fig:OPV_plots} and \ref{fig:photon_phases}.

We checked the resulting gamma-ray timing solutions using the radio ToAs, which were generally in agreement, but revealed small residual trends (on the order of 1\% to 10\% of a rotation) that were consistent with residual variations in the orbital phase that are below the level of precision provided by the gamma-ray timing solution. Owing to these remaining residuals we have not attempted to phase-align the radio and gamma-ray pulse profiles. A modification of the above timing procedure to jointly fit radio ToAs and gamma-ray photon arrival times would likely resolve this. 

\subsection{Gamma-ray eclipses}
\citet{Clark2023_eclipses} detected evidence in the \textit{Fermi}-LAT data for gamma-ray eclipses from both PSRs~J0838$-$2827 and J2333$-$5526, but the eclipse properties were uncertain since no pulsar timing ephemeris was available to precisely determine the orbital period and phase parameters. No gamma-ray eclipse was detected from PSR~J0955$-$3947. We confirm the detected gamma-ray eclipses in PSRs~J0838$-$2827 and J2333$-$5526, and the non-detection for PSR~J0955$-$3947, now using the precise gamma-ray timing solutions derived above to compute the orbital phases and to re-weight the photons according to their measured spin phase and best-fitting pulse profile template. For PSRs~J0838$-$2827 and PSR~J2333$-$5526, we find that the eclipse durations are 0.015 to 0.034 orbits, and 0.045 to 0.07 orbits, respectively (95\% confidence intervals). Using the now-known mass ratio, and assuming that the companion star fills its Roche lobe, and that there is no significant gamma-ray absorption from intra-binary material outside the Roche lobe, this sets a lower bound on the binary inclination angle of $i > 78^{\circ}$ for both pulsars. For PSR~J0955$-$3947, the eclipse non-detection implies an inclination $i < 76^{\circ}$ under the same assumptions.

These inclination constraints for PSRs J0955$-$3947 and J2333$-$5526 are also qualitatively consistent with the shapes of the gamma-ray pulse profiles, under the assumption that the pulsars achieved their observed spin rates through accretion from their companion stars such that their current spin axis is aligned with the orbit. For PSR~J2333$-$5526, the two sharp peaks are very close to half a rotation apart, which is what one would expect from a gamma-ray pulsar viewed from close to the spin/binary equator \citep[e.g.][]{Watters2008}. The narrower separation between peaks in the pulse profile for PSR~J0955$-$3947 is similarly indicative of a more moderate viewing angle/inclination. Future modelling of the phase-aligned radio and gamma-ray pulse profiles \citep[e.g.][]{Johnson2014,Corongiu2021+J2039} may provide a quantitative test of these constraints, and also for PSR~J0838$-$2827 whose broader pulse profile is somewhat more difficult to interpret. 

\subsection{Radio eclipses}
\label{s:eclipses}
We have detected radio eclipses in the data from PSRs J0838$-$2827 and J0955$-$3947. The orbital phase coverage of the various observations for all three new MSPs is shown in Figure \ref{fig:phase_cover} so that their eclipses can be contextualised.

Radio eclipsing is often seen in spider systems as gas outflowing from the companion star can block radio pulsations for extended parts of the orbit, generally near superior conjunction of the pulsar. Synchrotron absorption is established as the primary eclipse mechanism acting in most systems \citep{Polzin_2018} and whether the pulsar disappears completely or not, depends on the radio frequency compared to the characteristic plasma frequency of the eclipsing medium \citep{Thompson_1994}. Additionally, the intervening plasma causes a local increase in the dispersion measure, which is seen as a delay curving the pulse trail.

The \textit{L}-band discovery detection of PSR~J0838$-$2827 in February 2021 displays an eclipse ingress around 20\,min into the 1\,hr observation followed by a total eclipse (see Figure \ref{fig:J0838_eclipse}). Another observation at UHF in June 2021 shows no sign of eclipse in another 1\,hr long observation covering a similar range of the orbit. Since eclipses tend to last longer at lower frequencies, this clearly indicates that the medium responsible for the absorption is highly variable over time. The orbital range in which the eclipse took place, $0.7-0.8$, is also unusual as this corresponds to the inferior conjunction of the pulsar (i.e. half an orbit away from where it normally happens) while NRT and Parkes timing observations show a typical eclipse around phase $0.4$. An archival GBT observation covering orbital phases $0.5-0.6$ also displayed a complete eclipse for part of the observation. Finally, Parkes archival data from 2017 shows a total eclipse for half of the observation in phases $0.25-0.40$ (see Figure \ref{fig:old_parkes}). For non-detection observations, we ran \texttt{Spider\_twister} to check that the disappearances are due to a high orbital frequency derivative of the system with an inadequate preliminary ephemeris or actual eclipses. The latter seems to be the case as \texttt{Spider\_twister} could not recover a pulsation from those observations.

For PSR~J0955$-$3947, two MeerKAT observations display several short, transient eclipses lasting around 40\,s near orbital phase 0.1 (see Figure \ref{fig:J0955_mini_eclipse}). These broadband flux variations cannot be explained by interstellar scintillation, which for this pulsar would have a scintillation bandwidth of around 0.2 MHz according to NE2001 \citep{ne2001}. From the follow-up campaign, four of our Parkes observations starting close to orbital phase 0.4 show the pulsar coming out of the eclipse. While by design we have avoided observing near superior conjunction of the pulsar (phase 0.25), it appears quite likely that PSR~J0955$-$3947 experiences ``classic'' eclipses between phases 0.1 and 0.4, with potentially short eclipses on the ingress side.

No eclipses were detected in the case of PSR~J2333$-$5526. This is partly due to the fact that our observing scheme was successful in avoiding the superior conjunction of the pulsar (around phase 0.25) where regular eclipses normally take place. We can not rule out that eclipses can occur in this range; dedicated observations would be needed to investigate this. Note that this pulsar tends to exhibit very strong scintillation, which is expected given its low DM. We, therefore, conclude that scintillation is the most likely cause for the non-detection in some of the Murriyang Parkes observations, also owing to the fact that the telescope is less sensitive than MeerKAT.

\section{Discussion}
\label{S:discussion}
In Section~\ref{S:result}, we presented the discovery and timing of three new millisecond pulsars in our targeted MeerKAT searches of six high-confidence redback candidates. In this Section, we discuss possible explanations for the lack of detections from the remaining three candidates and discuss the new insights into our discoveries that our pulsation detections and timing solutions provide. 

\subsection{Candidates without pulsation detections}
We did not detect any convincing candidate pulsar signals from 4FGL J0523.3$-$2527, J0940.3$-$7610 and J1120.0$-$2204. Possibly, our non-detections are simply due to intrinsic faintness. While these candidates are all found in relatively bright gamma-ray sources \citep{4FGLDR4}, pulsar radio fluxes are almost entirely uncorrelated with gamma-ray fluxes \citep{Smith2023+3PC}, and so they may yet host very faint radio pulsars. A handful of ``radio-quiet'' gamma-ray MSPs are also known, thought to be due to emission and viewing geometries being such that their radio beams do not sweep across our line of sight \citep{Clark_2017,Smith2023+3PC}. To estimate a nominal flux density upper limit for our observations, we use the pulsar radiometer equation \citep{PSRHandbook},
\begin{equation}
S_{\rm min} = \frac{ {\rm S/N} \, (T_{\rm sys}+T_{\rm sky}) }{ \beta \, G \sqrt{n_{\rm pol} t_{\rm obs} {\rm BW}} } \sqrt{\frac{W}{P-W}}  \,,
\label{E:radiometerpulsar}
\end{equation}
where ${\rm S/N} = 9$ is the signal-to-noise ratio required for a confident detection, $T_{\rm sys} = 22.5\,$K is the system temperature, $T_{\rm sky}$ is the sky temperature taken from the 408 MHz all-sky map \citep{Haslam_1982,Remazeilles_2015} scaling to the central frequency of 1284\,MHz assuming a spectral index of --2.6, $G = 2.8 \, \rm K\,Jy^{-1}$ is the gain, $n_{\rm pol} = 2$ is the number of polarisations, $BW = 700$ and 500\,MHz is the effective usable bandwidth at $L$-band and UHF, $t_{\rm obs}$ is the observing time, $w = 0.15$ is the assumed fractional pulse width, and $\beta \approx 0.7$ conservatively accounts for various sources of sensitivity losses in the searching process (digitisation, beamforming efficiency, incoherent harmonic summing, FFT scalloping, acceleration and DM smearing, etc.). For the 15\,min, 30\,min and 60\,min segments that we searched, we find conservative flux density upper limits of around $40\,\mu$Jy, $30\,\mu$Jy and $20\,\mu$Jy (\textit{L}-band) and $60\,\mu$Jy, $45\,\mu$Jy and $30\,\mu$Jy (UHF), respectively. For an assumed distance of 2.2\,kpc (for 4FGL J0523.3$-$2527, the most distant of the unconfirmed candidates) a putative pulsar would have to have a pseudo-luminosity $L_{\rm 1400} = S_{\rm 1400} d^2$ fainter than 95\% of known MSPs in the ATNF Pulsar Catalogue \citep{psrcat} to escape detection in our 30\,min segments. 

Perhaps the most likely explanation for these non-detections is that their pulsations are nearly always rendered undetectable by eclipsing intra-binary material. While we attempted to mitigate this effect by observing these systems when the pulsar was in front of the companion  (at orbital phases between $0.5$--$1.0$) according to the published optical ephemerides, transient eclipses are often observed in other redbacks at unexpected orbital phases \citep[e.g.][]{Roy_2014,Deneva2016+J1048}, even around the pulsar's inferior conjunction \citep{Kudale2020+J1227}, implying that material may sometimes entirely enshroud the binary systems. The recent discovery of PSR~J0212$+$5320 \citep{Perez_2023}, which was only detected in two out of five GBT observations targeting an optical redback candidate, is a particularly relevant example for our study. Our MeerKAT timing observations of PSR~J0955$-$3947 (Figure~\ref{fig:J0955_mini_eclipse}) also exhibit short (lasting around 40\,s) transient eclipses at an orbital phase close to the eclipse ingress, but still outside of the eclipsing region.
 
Among the sources with no pulsar identified, flaring activity has been reported in the optical light curve of 4FGL J0523.3$-$2527, which could be a result of magnetic field reconnection events due to intra-binary material interacting with the shocked pulsar wind \citep{Sironi_2011,Halpern_2022}. While our repeated observations were intended to increase our chances of detection despite the presence of transient eclipses and interstellar scintillation \citep[e.g.][]{Camilo_2015}, at least one spider binary is detectable less than 10\% of the time \citep{Ray2013+J1311}, and so further observations of these high-confidence targets remain warranted. 

Indeed, our discovery of PSR~J0838$-$2527 provides another extreme case of redback elusiveness. The discovery observation (Figure~\ref{fig:J0838_eclipse}) shows the signal disappearing due to an apparent eclipse at an orbital phase of 0.7, and following another hour-long detection at UHF in June 2021, this pulsar remained undetectable until August 2023, in spite of multiple monitoring observations. A possible low-significance detection with NRT was made half way through that period, but would have been dismissed without prior knowledge of the timing ephemeris. Observations of transient X-ray flares and optical emission lines from this system suggest that substantial material overflows the companion's Roche lobe \citep{Halpern_2017}, providing a source of eclipsing material.

Another tantalising explanation for our non-detections is that this pulsar may be a transitional millisecond pulsar (tMSP) that switches from a radio millisecond pulsar state to a low-mass X-ray binary (LMXB) state, as proposed by \citet{Rea_2017} based on X-ray flaring activity. However, the 14-year 4FGL-DR4 light curve shows no evidence of gamma-ray flux variability, which has been observed in the two known Galactic tMSP transitions that have occurred during the \textit{Fermi} mission \citep{Stappers_2014,Johnson2015+J1227}. Hence, while it might not classify as a full transitional MSP, PSR~J0838$-$2527 might be a less extreme version in which the pulsar vanishes for extended periods due to material lying in the system but no accretion disc forms.

\subsection{Orbital period variations}
In Section~\ref{s:gamma-ray}, we applied the gamma-ray timing techniques introduced in \citet{Clark2021+J2039} to model the redback orbital period variations as stochastic Gaussian processes. Since this procedure has not yet been applied to the full population of \textit{Fermi}-LAT redbacks, we do not have a lot of context in which to discuss these results. Nevertheless, the results for these three pulsars already reveal some interesting differences with respect to each other, and with PSR~J2039$-$5617, discussed in \citet{Clark2021+J2039}. 

The traditional explanation for the orbital period variations commonly seen in spider binaries is the Applegate mechanism \citep{Applegate1992}, in which variations in gravitational quadrupole moment of the companion star, perhaps driven by magnetic activity, couple with the binary orbital period \citep[e.g.][]{Applegate1994+B1957,Lazaridis2011+J2051,Pletsch2015+J2339}. Under this model, the change in the companion quadrupole moment ($Q$) is directly related to the change in the orbital period, via $\Delta Q/Q \propto \Delta P_{\rm b}/P_{\rm b}$ \citep{Voisin2020}, which we can estimate from the derivatives of the Gaussian process described above, with $\Delta P_{\rm b}/P_{\rm b} \lesssim 10^{-6}$ for all three pulsars. Following the equations in \citet{Voisin2020}, and assuming an apsidal motion constant $k_2 \gtrsim 10^{-3}$, we find that $\Delta Q / Q \lesssim 2\%$ for each case. 

The spectral indices of the power spectra of the orbital phase variations are also of potential interest, as this parameter encodes information about the mechanism driving the orbital period changes. For PSR~J0955$-$3947, we find that the orbital phase variations are well described by a pure power-law process, with a length scale, $\ell$, likely to be longer than our $15$-year data set, and with Mat\'{e}rn degree $\nu=1.4 \pm 0.1$. This is strikingly close to, and statistically compatible with $\nu = 1.5$ (equivalently, $\gamma = -4$), which is the expected spectral index for a random-walk process in the orbital period. This therefore implies that the star's quadrupole moment undergoes a random walk. In contrast, for PSR~J2333$-$5526, a smoothly broken power law is preferred, with characteristic length scale $300 \, {\rm d} < \ell < 800 \, {\rm d}$, well within the frequencies probed by our data, breaking to a steep power-law process (with $\gamma > 5$) at higher frequencies. The steeper index implies that these changes are more gradual than in PSR~J0955$-$3947, and that there are time-varying torques (e.g. mass loss, asynchronous rotation) acting on the companion star.

In this respect, PSR~J2333$-$5526 is much more similar to PSR~J2039$-$5617, which has a similar characteristic length scale, and a steeply-decreasing spectrum at higher frequencies \citep{Clark2021+J2039}.  The picture is less clear for PSR~J0838$-$2827. Its lower spin frequency and smaller semi-major axis compared to the other two pulsars increase the uncertainty on the orbital phase, leading to much greater uncertainty in the hyperparameters. Here the power spectrum could either be a pure, shallow power-law like J0955$-$3947, or could break to a steeper power low at high frequencies, like J2333$-$5526. 

The orbital period variations seen in PSRs J0955$-$3947 and J2333$-$5526 have a larger amplitude, and are more complicated, than those that have previously been subjected to gamma-ray timing studies (PSR~J2339$-$0533 in \citealt{Pletsch2015+J2339} and PSR~J2039$-$5617 in \citealt{Clark2021+J2039}). These pulsars required substantial improvements in our methodology to reach a satisfactory timing solution. Aside from allowing the orbital period variations in redbacks to be studied more quantitatively, our new methods also provide a possible path for bright gamma-ray redbacks to contribute to the Gamma-ray Pulsar Timing Array \citep[$\gamma$PTA,][]{GPTA}, several of which have been excluded until now due to their complicated behaviour. Both PSRs~J0955$-$3947 and J2333$-$5526 have log-likelihoods above the threshold for inclusion in the $\gamma$PTA. 

For PSR~J0838$-$2827, we required four spin frequency derivatives to model the phase evolution over the \textit{Fermi}-LAT mission. These terms are unusual for a gamma-ray MSP, where long-term red noise is all but unmeasurable in gamma-ray data for most MSPs \citep{GPTA}. In one other black-widow binary, PSR~J1555$-$2908, the presence of significant higher order spin frequency variations has been interpreted as possible evidence for a third, planetary mass object orbiting the inner binary system \citep{Nieder2022+J1555}. Further timing analyses of this pulsar as the \textit{Fermi}-LAT mission continues may determine whether or not this timing noise is intrinsic to the pulsar. 

\subsection{Distances and energetics}
The detection of radio pulsations from these MSPs provides knowledge of their DMs, which can be used to infer distances based on models of the Galactic electron density content. These can be compared to distances estimates obtained from \textit{Gaia} parallax measurements (or upper limits), and constraints from previous optical modelling. 

For PSRs~J0838$-$2827 and J0955$-$3947 there is significant tension between the DM-inferred distances ($d=410$\,pc and $d=490$\,pc, respectively, according to YMW16) and the lower bounds from \textit{Gaia} DR3 parallax ($\varpi = 0.43\pm0.40$\,mas and $\varpi = 0.27\pm0.13$\,mas, respectively, where the 2$\sigma$ upper limits correspond to minimum distances of $d=800$\,pc and $d=1.9$\,kpc, respectively). Upon closer inspection, these pulsars lie close to the edge of the Gum Nebula, perhaps indicating that the shell-shaped excess electron density added to the YMW16 model to account for this feature may be overestimated in these directions. The NE2001 model predicts larger distances for these pulsars, $d=560$\,pc and $d=3.3$\,kpc, respectively, compatible with the parallax range for PSR~J0955$-$3947 but still not with that for J0838$-$2827. For PSR~J2333$-$5526, the \textit{Gaia} DR3 parallax ($\varpi = -0.37\pm0.52$, for $d > 1.5$\,kpc) all but rules out the NE2001 distance ($d=860$\,pc), but the YMW16 distance $d=2.5$\,kpc is plausible. For further calculations, we adopt nominal distances of: $d=1.7$\,kpc for PSR~J0838$-$2827 (the maximum distance found by \citealt{Halpern_2017+0838MSP} to be compatible with the observed optical flux); $d = 3.5$\,kpc for PSR~J0955$-$3947 (the distance obtained by \citealt{Koljonen2023+Gaia} using the \textit{Gaia} DR3 parallax); and $d=3.1$\,kpc for PSR~J2333$-$5526 (from optical modelling by \citealt{Swihart_2020}).

The gamma-ray timing solutions, constrained by \textit{Gaia} DR3 astrometry, also provide precise values for proper motion that can be used to Doppler-correct the observed spin-down rates, and therefore the spin-down energy budget, of these pulsars for the radial acceleration induced by their transverse motion, known as the ``Shklosvkii effect'' \citep{Shklovskii}, and for their acceleration in the Galactic potential. For PSRs~J0955$-$3947 and J2333$-$5526 these corrections are small, totalling less than $10$\% of the observed spin-down rate, but for PSR~J0838$-$2827 the Shklovskii component is larger, accounting for more than $20$\% of the observed $\dot{f}$ at $d=1.7$\,kpc. These spin-down corrections are included in the derived parameter values given in Table~\ref{T:timing}. 

At our nominal distances, using the above Doppler corrections an assuming isotropic time-averaged emission, the observed gamma-ray fluxes, $G_{\gamma}$, from the 4FGL-DR4 catalogue \citep{4FGLDR4} correspond to gamma-ray efficiencies $\eta = 4 \pi d^2 G_{\gamma} / \dot{E}$ of $37\%$ (PSR~J0838$-$2827), $7\%$ (PSR~J0955$-$3947) and $14\%$ (PSR~J2333$-$5526), which are typical for gamma-ray MSPs \citep{Smith2023+3PC}. 

\citet{Li_2018} and \citet{Swihart_2020} report ``day'' temperatures of $8000$\,K and $5700$\,K and ``night'' temperatures of $5700$\,K and $4400$\,K for PSRs~J0955$-$3947 and J2333$-$5526, respectively. These can also be compared to the spin-down luminosity by computing the irradiation efficiency, $\epsilon = 4 \pi A_1^2 (1 + q)^2 \sigma_{\rm SB} (T_{\rm day}^4 - T_{\rm night}^4) / \dot{E}$ \citep{Breton_2013}, where $\sigma_{\rm SB}$ is the Stefan-Boltzmann constant. We find $\epsilon \approx 40\%$ for both pulsars, which is typical for irradiated spider binaries \citep[e.g.][]{MataSanchez2023}

\subsection{Mass measurement}
\label{S:mass}
Our timing solutions provide precise values for the pulsars' projected semi-major axes, $A_1$, which were the missing pieces for pulsar mass measurements for these systems. 

As discussed above, the presence/absence of gamma-ray eclipses constrain the binary inclination angle, $i$, of PSRs~J0838$-$2827, J0955$-$4947 and J2333$-$5526, to be $i > 78\degr$, $i < 76\degr$ and $i > 78\degr$, respectively. 

The pulsar mass, $M_{\rm psr}$, can be estimated from these parameters via the binary mass function,
\begin{equation}
M_{\rm psr} \sin^3{i} = P_{\rm b} K_{2}^3 \frac{(1 + K_{1}/K_{2})^2}{2 \pi G}\,,
\label{E:massfunction}
\end{equation}
where $P_{\rm b}$ is the orbital period, $K_{1} = 2\pi A_1/P_{\rm b}$ is the pulsar's projected radial velocity, $K_{2}$ is the companion radial velocity amplitude and $G$ is the gravitational constant. The companion radial velocities, reported in Table~\ref{T:timing}, were obtained via optical spectroscopy taken when these binary systems were first identified as candidate redbacks\footnote{We discovered that the orbital phase from the gamma-ray timing ephemeris for PSR~J2333$-$5526 disagreed with that measured from optical spectroscopy \citet{Swihart_2020}. This has recently been found to be due to flexure in the SOAR/Goodman spectrograph \citep{Dodge2024+1910}. Correcting for this effect gives a revised radial velocity amplitude of $K_2 = 362.5\pm 7.6$\,km s$^{-1}$ (Strader, J., private communication).}.

We estimated pulsar masses using a Monte-Carlo technique, in which we generate 50,000 random realisations, with values for $K_2$ and $A_1$ drawn from normal distributions according to our timing solutions, and with $\cos i$ drawn from a uniform distribution (representing an isotropic distribution for the binary orbital axis), discarding values lying outside the constraints from the gamma-ray eclipse observations (e.g. requiring $i > 78\degr$ for PSR~J0838$-$2827). The results of these calculations are presented in Table~\ref{T:timing}. These values span a large range, from $M_{\rm psr} = 0.93\pm0.14 {\rm M}_{\odot}$ for PSR~J0838$-$2527, to $M_{\rm psr} = 2.06\pm0.13 {\rm M}_{\odot}$ for PSR~J2333$-$5526. 

However, one significant source of systematic uncertainty in these values remains, which is the ``centre-of-light'' adjustment to $K_2$, required to correct the observed radial velocity curves to account for the effects of the large difference in surface temperature between the irradiated and un-irradiated faces of the companion star. These corrections are difficult to estimate, as they vary between different spectral lines, and depend strongly on the temperature pattern on the surface of the companion star. \citet{Swihart_2020} account for this in an approximate manner for PSR~J2333$-$5526 by producing a model of the companion star surface temperatures, and using an effective temperature--equivalent width relation for the adopted Mg\,b triplet to produce a ``corrected'' model radial velocity curve, which reduces the inferred $K_2$. Adopting this value, we find a lower $M_{\rm psr} = 1.70 \pm 0.06 {\rm M}_{\odot}$. However, this is at odds with the picture found by \citet{Kandel2020} for the similarly-irradiated PSR~J2215$+$5135, in which the Mg\,b lines only slightly \emph{under}-estimate the centre-of-mass velocity, suggesting that the original uncorrected value may be closer to reality. \citet{Dodge2024+1910} find almost no centre-of-light correction is required to the velocity of the Mg\,b line in a similarly irradiated redback, PSR~J1910$-$5320, with an uncertainty of around 4\% on the correction, suggesting a systematic uncertainty here of around 0.2--0.3\,M$_{\odot}$ for PSR~J2333$-$5526. 

Estimates for the centre-of-light corrections have not been made for either PSR~J0838$-$2527 or J0955$-$3947. For both pulsars, radial velocities were predominantly measured from the Ca II triplet \citep{Halpern_2017+0838MSP,Li_2018}, whose equivalent width varies little with temperature \citep{Mallik1997} and which will therefore track a position close to the heated face of the companion star which contributes most strongly to the observed spectra. The true centre-of-mass radial velocity amplitude will therefore be larger than observed, by up to around 20\% \citep[e.g.][]{Romani2011+J2339}, meaning our pulsar mass estimates are better interpreted as lower bounds for these pulsars, and the true mass could be up to 60\% larger.   

A full modelling treatment of the optical light curves and spectra (or radial velocities) \citep[e.g.][]{Linares_2018,Kandel2020,Kennedy2022+J1555} will be necessary to turn our lower limits into more precise pulsar mass estimates for these systems. For PSR~J0838$-$2527 in particular, such an effort will likely be complicated by the strong asymmetry and variability observed by \citet{Cho2018}. Nevertheless, our results are consistent with the picture from \citet{Strader_2019} that redbacks tend to contain rather heavy neutron stars, and in particular the mass of PSR~J2333$-$5526 is likely to lie towards the heavier end of this distribution.

\section{Conclusion and Future work}
\label{S:conclusion}
We have presented the results from TRAPUM’s deep survey for six redback candidates using the MeerKAT radio telescope. We found pulsations from MSPs in three of these systems, thus confirming their nature as redbacks. We detected PSRs~J0955$-$3947 and J2333$-$5526 in most of our search and timing observations. The data we collected enabled us to establish phase-connected timing solutions, which enabled the detection of gamma-ray pulsations and a full 15-year gamma-ray timing solution accounting for significant stochastic variability in the orbital periods, consistent with gravitational quadrupole moment variations in the companion star.

On the contrary, PSR~J0838-2827 remained undetected for two years after our discovery, except for a single faint detection with NRT, until August 2023 when we started to re-detect this source with MeerKAT and Parkes. With these new detections, we could find a local phase-connected timing solution, which enabled gamma-ray timing and finding gamma-ray pulsation. We should note that PSR~J0838$-$2827 has been reported as a possible transitional millisecond pulsar candidate \citep{Rea_2017}, but gamma-ray pulsations are detectable throughout 15 years of monitoring with the \textit{Fermi}-LAT, with no apparent flux variations or transitions.

We used the additional orbital information derived from these discoveries to update the estimated pulsar masses in these three systems. Accordingly, PSRs~J0838$-$2827, J0955$-$3947, and PSR~J2333$-$5526 have minimum pulsar masses of $0.93 \pm 0.14 \textrm{M}_{\odot}$, $1.33 \pm 0.05 \textrm{M}_{\odot}$, and $2.06 \pm 0.13 \textrm{M}_{\odot}$.

We encourage further monitoring of PSR~J0838$-$2827, as its elusive behaviour, flaring X-ray activity and optical emission features might all point towards future transitional behaviour between pulsar and X-ray binary-like states. Likewise, we suggest that our inability to detect pulsations in the other three candidates is likely caused by a combination of heavy eclipses and potentially faint pulsations. Further observations might eventually unravel these pulsars.

Our discovery of three redback MSPs from six gamma-ray sources that had been targeted repeatedly by other telescopes, without success, reinforces the discovery potential of MeerKAT MSP searches. A number of new spider candidates have been discovered in \textit{Fermi}-LAT sources since we began our survey: 4FGL J0336.0+7502 \citep{Li_2021}, 4FGL J0935.3+0901 \citep{Halpern_2022+0935}, 4FGL J1408.6$-$2917 \citep{Swihart_2022+1408} contain candidate black widows while 4FGL J1702.7$-$5655 \citep{Corbet_2022} and 4FGL J2054.2+6904 \citep{Karpova_2023} host candidate redbacks. We have begun performing similar searches of the more southerly of these.

\section*{Acknowledgements}
The MeerKAT telescope is operated by the South African Radio Astronomy Observatory (SARAO), which is a facility of the National Research Foundation, an agency of the Department of Science and Innovation. We thank staff at SARAO for their help with observations and commissioning. TRAPUM observations used the FBFUSE and APSUSE computing clusters for data acquisition, storage and analysis. These clusters were funded and installed by the Max-Planck-Institut f\"{u}r Radioastronomie (MPIfR) and the Max-Planck-Gesellschaft. The National Radio Astronomy Observatory is a facility of the National Science Foundation operated under cooperative agreement by Associated Universities, Inc. The Parkes radio telescope is part of the Australia Telescope National Facility (https://ror.org/05qajvd42) which is funded by the Australian Government for operation as a National Facility managed by CSIRO. We acknowledge the Wiradjuri people as the traditional owners of the Observatory site. The Nan\c cay Radio Observatory is operated by the Paris Observatory, associated with the French Centre National de la Recherche Scientifique (CNRS) and Universit\'{e} d'Orl\'{e}ans. It is partially supported by the Region Centre Val de Loire in France.

The \textit{Fermi} LAT Collaboration acknowledges generous ongoing support from a number of agencies and institutes that have supported both the development and the operation of the  LAT  as  well  as  scientific  data  analysis.  These  include  the National  Aeronautics  and Space   Administration   and   the   Department   of   Energy   in the   United   States,   the Commissariat \`{a} l'Energie Atomique and the Centre National de la Recherche Scientifique /  Institut  National  de  Physique  Nucl\'{e}aire et  de  Physique  des  Particules  in  France,  the Agenzia  Spaziale  Italiana and the Istituto  Nazionale  di  Fisica  Nucleare  in  Italy,  the Ministry  of  Education,  Culture, Sports,  Science  and  Technology  (MEXT),  High  Energy Accelerator  Research Organization  (KEK)  and  Japan  Aerospace  Exploration  Agency (JAXA)  in  Japan, and  the  K.  A.  Wallenberg  Foundation,  the  Swedish  Research  Council and the Swedish National Space Board in Sweden. 

This work has made use of data from the European Space Agency (ESA) mission {\it Gaia} (\url{https://www.cosmos.esa.int/gaia}), processed by the {\it Gaia} Data Processing and Analysis Consortium (DPAC, \url{https://www.cosmos.esa.int/web/gaia/dpac/consortium}). Funding for the DPAC has been provided by national institutions, in particular the institutions participating in the {\it Gaia} Multilateral Agreement.

T.T. is grateful to the National Astronomical Research Institute of Thailand (NARIT) for awarding a student scholarship. R.P.B. and C.J.C. acknowledge support from the European Research Council (ERC) under the European Union's Horizon 2020 research and innovation programme (grant agreement No. 715051; Spiders). M.B. and A.P. gratefully acknowledges financial support by the research grant 2022 “GCjewels” (P.I. Andrea Possenti) approved with the INAF Presidential Decree 30/2022. E.D.B., C.J.C., D.J.C., P.C.C.F., M.K., L.N., P.V.P., A.R. and V.V.K. acknowledge continuing valuable support from the Max-Planck Society.  B.W.S. acknowledges funding from the ERC under the European Union’s Horizon 2020 research and innovation programme (grant agreement No. 694745). S.M.R. is a CIFAR Fellow and is supported by the NSF Physics Frontiers Center awards 2020265. F.C. acknowledges financial support from the ‘Agence Nationale de la Recherche’, grant number ANR-19-CE310005-01 (PI: Francesca Calore). V.V.K. acknowledges financial support from the European Research Council (ERC) starting grant ,``COMPACT'' (Grant agreement number 101078094), under the European Union's Horizon Europe research and innovation programme. L.V. acknowledges financial support from the Dean’s Doctoral Scholar Award from the University of Manchester.

We would like to thank Philippe Bruel, Matthew Kerr and David Smith for helpful comments on the manuscript. We also appreciate Jay Strader for revisiting the optical spectroscopy for PSR~J2333$-$5526. T.T. thanks Oliver Dodge and Adipol Phosrisom for their general comments. C.J.C would like to thank Rutger van Haasteren for helpful discussion and advice regarding gamma-ray pulsar timing analyses. Lastly, we appreciate the valuable comments from the anonymous reviewer.

\section*{Data Availability}
TRAPUM data products are available upon reasonable request to the TRAPUM collaboration. The \textit{Fermi}-LAT data are available from the \textit{Fermi} Science Support Center (FSSC, \url{http://fermi.gsfc.nasa.gov/ssc}). Gamma-ray timing solutions are also available from the FSSC (\url{https://fermi.gsfc.nasa.gov/ssc/data/access/lat/ephems/}).



\bibliographystyle{mnras}
\bibliography{example} 




\appendix
\section{Observations}
\begin{table*}
	\caption{
Observations of three MSPs discovered in our survey. The table contains the epoch of the beginning of each observation and the orbital phase coverage. MS is a 1\,hr MeerKAT searching campaign, MT is a 5 min MeerKAT timing campaign, T is the timing campaign with Parkes (PKS) or Nan\c{c}ay NUPPI and Ar is the archival data.}
	\centering
    \renewcommand{\arraystretch}{1.0}
    \setlength\tabcolsep{1.0pt}
    \label{T:epoch}
    \renewcommand{\arraystretch}{1.0}
    \setlength\tabcolsep{4.0pt}
	\begin{tabular}{cccccccc}
		\hline
        \hline
        \multicolumn{8}{c}{PSR~J0838$-$2827}\\
        OBS ID & Epoch(MJD) & Band & Number of dishes & Orbital phase (start) & Orbital phase (stop) & Observing campaign&Detection\\
		\hline
1 & 57800.40416667 & PKS-L & 1 & 0.287 & 0.481 & Ar & Yes\\
2 & 57800.44710648 & PKS-L & 1 & 0.487 & 0.683 & Ar & No\\
3 & 57800.49004630 & PKS-L & 1 & 0.687 & 0.883 & Ar & No\\
4 & 57800.53368056 & PKS-L & 1 & 0.890 & 0.072 & Ar & No\\
5 & 57801.36817130 & PKS-L & 1 & 0.780 & 0.971 & Ar & No\\
6 & 57801.50590278 & PKS-L & 1 & 0.422 & 0.617 & Ar & Yes\\
7 & 57804.45648148 & PKS-L & 1 & 0.176 & 0.372 & Ar & No\\
8 & 57805.45763889 & PKS-L & 1 & 0.843 & 0.039 & Ar & No\\
9 & 57806.51944444 & PKS-L & 1 & 0.793 & 0.984 & Ar & No\\
10 & 58726.55043449 & GBT-S & 1 & 0.524 & 0.574 & Ar & Yes\\
11 & 59281.75961995 & L & 60 & 0.635 & 0.813 & MS & Yes\\
12 & 59383.66282510 & UHF & 56 & 0.658 & 0.836 & MS & Yes\\
13 & 59589.50384377 & PKS-UWL & 1 & 0.187 & 0.763 & T & No\\
14 & 59596.49794099 & PKS-UWL & 1 & 0.790 & 0.027 & T & No\\
15 & 59604.63619331 & PKS-UWL & 1 & 0.727 & 0.862 & T & No\\
16 & 59605.98107639 & Nan\c{c}ay & 1 & 0.996 & 0.168 & T & No\\
17 & 59623.53764007 & PKS-UWL & 1 & 0.836 & 0.005 & T & No\\
18 & 59631.42288312 & PKS-UWL & 1 & 0.593 & 0.785 & T & No\\
19 & 59646.73451911 & L & 60 & 0.968 & 0.987 & MT & No\\
20 & 59646.77284207 & L & 60 & 0.147 & 0.161 & MT & No\\
21 & 59649.50915627 & PKS-UWL & 1 & 0.902 & 0.072 & T & No\\
22 & 59674.54898266 & PKS-UWL & 1 & 0.626 & 0.818 & T & No\\
23 & 59681.83974899 & L & 60 & 0.612 & 0.627 & MT & No\\
24 & 59681.86276031 & L & 60 & 0.719 & 0.734 & MT & No\\
25 & 59681.88707102 & L & 60 & 0.832 & 0.847 & MT & No\\
26 & 59681.91860452 & L & 60 & 0.979 & 0.994 & MT & No\\
27 & 59682.78305556 & Nan\c{c}ay & 1 & 0.009 & 0.153 & T & No\\
28 & 59690.83589324 & UHF & 64 & 0.547 & 0.725 & MS & No\\
29 & 59692.74902778 & Nan\c{c}ay & 1 & 0.465 & 0.643 & T & No\\
30 & 59692.79413470 & L & 60 & 0.676 & 0.854 & MS & No\\
31 & 59708.45767479 & PKS-UWL & 1 & 0.691 & 0.883 & T & No\\
32 & 59735.63174769 & Nan\c{c}ay & 1 & 0.363 & 0.559 & T & Yes\\
33 & 59737.34126275 & PKS-UWL & 1 & 0.332 & 0.682 & T & No\\
34 & 59743.60405093 & Nan\c{c}ay & 1 & 0.526 & 0.699 & T & No\\
35 & 59748.60000000 & Nan\c{c}ay & 1 & 0.815 & 0.969 & T & No\\
36 & 59768.12908683 & PKS-UWL & 1 & 0.850 & 0.202 & T & No\\
37 & 59813.41109954 & Nan\c{c}ay & 1 & 0.932 & 0.114 & T & No\\
38 & 59870.88750118 & PKS-UWL & 1 & 0.859 & 0.051 & T & No\\
39 & 59885.21931713 & Nan\c{c}ay & 1 & 0.667 & 0.825 & T & No\\
40 & 59893.19733796 & Nan\c{c}ay & 1 & 0.857 & 0.015 & T & No\\
41 & 59894.85643636 & PKS-UWL & 1 & 0.590 & 0.782 & T & No\\
42 & 59918.64192248 & PKS-UWL & 1 & 0.467 & 0.658 & T & No\\
43 & 59921.11826389 & Nan\c{c}ay & 1 & 0.010 & 0.183 & T & No\\
44 & 59922.53582294 & Nan\c{c}ay & 1 & 0.618 & 0.809 & T & No\\
45 & 60015.85888889 & Nan\c{c}ay & 1 & 0.644 & 0.817 & T & No\\
46 & 60157.21147230 & UHF & 56 & 0.560 & 0.609 & MT & No\\
47 & 60157.23683573 & UHF & 56 & 0.679 & 0.727 & MT & No\\
48 & 60157.26226770 & UHF & 56 & 0.797 & 0.846 & MT & No\\
49 & 60157.28397596 & UHF & 56 & 0.898 & 0.963 & MT & No\\
50 & 60158.49916342 & UHF & 56 & 0.563 & 0.579 & MT & Yes\\
    \hline
    \hline
	\end{tabular}
\end{table*}

\begin{table*}
	\centering
    \renewcommand{\arraystretch}{1.0}
    \setlength\tabcolsep{1.0pt}
    \renewcommand{\arraystretch}{1.0}
    \setlength\tabcolsep{4.0pt}
    \contcaption{List of observations.}
	\begin{tabular}{cccccccc}
        \hline
		\hline
        \multicolumn{8}{c}{PSR~J0838$-$2827}\\
        OBS ID & Epoch(MJD) & Band & Number od dishes & Orbital phase (start) & Orbital phase (stop) & Observing campaign & Detection\\
		\hline
51 & 60158.52476101 & UHF & 56 & 0.682 & 0.715 & MT & No\\
52 & 60160.65954023 & UHF & 60 & 0.634 & 0.682 & MT & No\\
53 & 60164.31571931 & UHF & 56 & 0.677 & 0.725 & MT & No\\
54 & 60174.43260694 & UHF & 56 & 0.837 & 0.885 & MT & No\\
55 & 60188.31964138 & UHF & 64 & 0.571 & 0.620 & MT & Yes\\
56 & 60219.06098495 & UHF & 60 & 0.872 & 0.938 & MT & No\\
57 & 60231.44079977 & UHF & 60 & 0.581 & 0.645 & MT & Yes\\
58 & 60235.87834609 & PKS-UWL & 1 & 0.266 & 0.458 & T & Yes\\
59 & 60239.81629748 & PKS-UWL & 1 & 0.623 & 0.996 & T & Yes\\
60 & 60240.86894794 & PKS-UWL & 1 & 0.530 & 0.717 & T & Yes\\
61 & 60241.81483914 & PKS-UWL & 1 & 0.940 & 0.266 & T & Yes\\
62 & 60242.81622803 & PKS-UWL & 1 & 0.607 & 0.260 & T & No\\
63 & 60243.80943405 & PKS-UWL & 1 & 0.237 & 0.053 & T & Yes\\
        \hline
        \multicolumn{8}{c}{PSR~J0955$-$3947}\\
        OBS ID & Epoch(MJD) & Band & Number of dishes & Orbital phase (start) & Orbital phase (stop) & Observing campaign & Detection\\
		\hline
1 & 58228.56180556 & PKS-L & 1 & 0.509 & 0.615 & Ar & Yes\\
2 & 58230.59166667 & PKS-L & 1 & 0.750 & 0.776 & Ar & Yes\\
3 & 58264.17939815 & PKS-L & 1 & 0.467 & 0.534 & Ar & Yes\\
4 & 58354.95011574 & PKS-L & 1 & 0.818 & 0.924 & Ar & Yes\\
5 & 58894.49988426 & PKS-L & 1 & 0.826 & 0.973 & Ar & Yes\\
6 & 59281.80189143 & L & 60 & 0.761 & 0.867 & MS & Yes\\
7 & 59383.70457396 & UHF & 56 & 0.853 & 0.951 & MS & Yes\\
8 & 59536.31265743 & L & 56 & 0.856 & 0.864 & MT & Yes\\
9 & 59536.35712501 & L & 56 & 0.971 & 0.978 & MT & Yes\\
10 & 59536.39231600 & L & 56 & 0.062 & 0.064 & MT & Yes\\
11 & 59536.39402915 & L & 56 & 0.066 & 0.068 & MT & Yes\\
12 & 59590.52284295 & PKS-UWL & 1 & 0.816 & 0.084 & T & Yes\\
13 & 59592.51183922 & PKS-UWL & 1 & 0.951 & 0.072 & T & Yes\\
14 & 59596.56176176 & PKS-UWL & 1 & 0.407 & 0.561 & T & Yes\\
15 & 59604.50972760 & PKS-UWL & 1 & 0.927 & 0.008 & T & Yes\\
16 & 59623.49592920 & PKS-UWL & 1 & 0.945 & 0.040 & T & Yes\\
17 & 59631.46848707 & PKS-UWL & 1 & 0.529 & 0.623 & T & Yes\\
18 & 59646.00292362 & L & 60 & 0.054 & 0.062 & MT & Yes\\
19 & 59646.08950694 & L & 60 & 0.277 & 0.285 & MT & No\\
20 & 59646.13135 & L & 60 & 0.386 & 0.394 & MT & No\\
21 & 59646.73866767 & L & 60 & 0.953 & 0.960 & MT & Yes\\
22 & 59646.77722474 & L & 60 & 0.053 & 0.058 & MT & Yes\\
23 & 59651.38246858 & PKS-UWL & 1 & 0.943 & 0.037 & T & Yes\\
24 & 59666.37406578 & PKS-UWL & 1 & 0.648 & 0.742 & T & Yes\\
25 & 59674.43511797 & PKS-UWL & 1 & 0.460 & 0.514 & T & Yes\\
26 & 59681.86624246 & L & 60 & 0.646 & 0.654 & MT & Yes\\
27 & 59681.89055962 & L & 60 & 0.708 & 0.717 & MT & Yes\\
28 & 59681.92215685 & L & 60 & 0.790 & 0.798 & MT & Yes\\
29 & 59681.95008508 & L & 60 & 0.862 & 0.870 & MT & Yes\\
30 & 59681.97825638 & L & 60 & 0.935 & 0.943 & MT & Yes\\
31 & 59698.46922625 & PKS-UWL & 1 & 0.511 & 0.586 & PT & Yes\\
32 & 59708.50078823 & PKS-UWL & 1 & 0.411 & 0.499 & PT & Yes\\
33 & 59738.37969079 & PKS-UWL & 1 & 0.552 & 0.646 & PT & Yes\\
34 & 59715.46715397 & PKS-UWL & 1 & 0.396 & 0.491 & PT & Yes\\
35 & 59768.13297573 & PKS-UWL & 1 & 0.369 & 0.520 & PT & Yes\\
36 & 59820.92296896 & PKS-UWL & 1 & 0.662 & 0.754 & PT & Yes\\
37 & 59846.04055885 & PKS-UWL & 1 & 0.510 & 0.605 & PT & Yes\\
38 & 59690.87758498 & UHF & 64 & 0.911 & 0.010 & MS & Yes\\
39 & 59692.83553331 & L & 60 & 0.966 & 0.065 & MS & No\\
    \hline
    \hline
	\end{tabular}
\end{table*}

\begin{table*}
	\centering
    \renewcommand{\arraystretch}{1.0}
    \setlength\tabcolsep{1.0pt}
    \renewcommand{\arraystretch}{1.0}
    \setlength\tabcolsep{4.0pt}
    \contcaption{List of observations.}
	\begin{tabular}{cccccccc}
		\hline
        \hline
        \multicolumn{8}{c}{PSR~J2333$-$5526}\\
        OBS ID & Epoch(MJD) & Band & Number of dishes & Orbital phase (start) & Orbital phase (stop) & Observing campaign & Detection\\
        \hline
1 & 57109.20520833 & PKS-L & 1 & 0.876 & 0.019 & Ar & No\\
2 & 57135.90474537 & PKS-L & 1 & 0.697 & 0.819 & Ar & No\\
3 & 57144.12280093 & PKS-L & 1 & 0.267 & 0.410 & Ar & No\\
4 & 57145.10266204 & PKS-L & 1 & 0.674 & 0.816 & Ar & No\\
5 & 57157.12905093 & PKS-L & 1 & 0.484 & 0.553 & Ar & No\\
6 & 57204.73506944 & PKS-L & 1 & 0.986 & 0.073 & Ar & No\\
7 & 57868.19013889 & PKS-L & 1 & 0.490 & 0.514 & Ar & No\\
8 & 58424.50358796 & PKS-L & 1 & 0.515 & 0.559 & Ar & No\\
9 & 58424.51770833 & PKS-L & 1 & 0.564 & 0.707 & Ar & No\\
10 & 58578.21901620 & PKS-UWL & 1 & 0.907 & 0.008 & Ar & Yes\\
11 & 58579.22269676 & PKS-UWL & 1 & 0.397 & 0.487 & Ar & No\\
12 & 58580.18784722 & PKS-UWL & 1 & 0.752 & 0.894 & Ar & No\\
13 & 58586.09050926 & PKS-L & 1 & 0.273 & 0.415 & Ar & No\\
14 & 58586.09041667 & PKS-UWL & 1 & 0.272 & 0.415 & Ar & Yes\\
15 & 59314.51063650 & L & 60 &0.627 & 0.763 & MS & Yes\\
16 & 59388.15962432 & UHF & 60 & 0.668 & 0.807 & MS & Yes\\
17 & 59500.56993172 & PKS-UWL & 1 & 0.463 & 0.934 & T & No\\
18 & 59696.85988437 & PKS-UWL & 1 & 0.866 & 0.958 & T & No\\
19 & 59635.25228129 & L & 60 & 0.687 & 0.696 & MT & Yes\\
20 & 59635.31725745 & L & 60 & 0.913 & 0.924 & MT & No\\
21 & 59674.16373963 & L & 64 & 0.962 & 0.971 & MT & Yes\\
22 & 59675.18489704 & L & 60 & 0.513 & 0.521 & MT & Yes\\
23 & 59675.23327667 & L & 60 & 0.681 & 0.690 & MT & Yes\\
24 & 59692.53882956 & L & 60 & 0.844 & 0.862 & MT & Yes\\
25 & 59692.50401411 & UHF & 60 & 0.722 & 0.862 & MS & Yes\\
26 & 59698.53888685 & L & 60 & 0.703 & 0.721 & MT & Yes\\
27 & 59698.50415936 & L & 60 & 0.582 & 0.721 & MS & Yes\\
28 & 59708.83588079 & PKS-UWL & 1 & 0.500 & 0.635 & T & Yes\\
29 & 59715.71597559 & PKS-UWL & 1 & 0.419 & 0.537 & T & Yes\\
30 & 59725.02419100 & L & 60 & 0.779 & 0.788 & MT & Yes\\
31 & 59726.11967709 & L & 60 & 0.587 & 0.599 & MT & Yes\\
32 & 59727.04198033 & L & 60 & 0.794 & 0.806 & MT & Yes\\
33 & 59728.13519794 & L & 60 & 0.594 & 0.603 & MT & Yes\\
34 & 59738.77092930 & PKS-UWL & 1 & 0.570 & 0.772 & T & Yes\\
35 & 59759.88238091 & UHF & 60 & 0.964 & 0.975 & MT & Yes\\
36 & 59760.16259956 & UHF & 60 & 0.938 & 0.947 & MT & Yes\\
37 & 59760.93386099 & UHF & 60 & 0.619 & 0.631 & MT & Yes\\
38 & 59762.92942827 & UHF & 60 & 0.557 & 0.566 & MT & Yes\\
39 & 59766.68657524 & PKS-UWL & 1 & 0.618 & 0.905 & T & Yes\\
40 & 59768.64775579 & PKS-UWL & 1 & 0.436 & 0.647 & T & Yes\\
41 & 59768.96579981 & L & 60 & 0.542 & 0.551 & MT & Yes\\
42 & 59793.47894792 & PKS-UWL & 1 & 0.762 & 0.998 & T & Yes\\
43 & 59797.80471181 & L & 60 & 0.801 & 0.813 & T & No\\
44 & 59820.73430672 & PKS-UWL & 1 & 0.516 & 0.802 & T & Yes\\
45 & 59828.75854284 & L & 56 & 0.412 & 0.424 & MT & No\\
46 & 59828.79346181 & L & 56 & 0.533 & 0.545 & MT & No\\
47 & 59830.47783681 & PKS-UWL & 1 & 0.389 & 0.616 & T & Yes\\
48 & 59851.65143638 & L & 60 & 0.999 & 0.008 & MT & Yes\\
49 & 59865.63141320 & PKS-UWL & 1 & 0.600 & 0.954 & T & No\\
50 & 59957.40894792 & PKS-UWL & 1 & 0.665 & 0.850 & T & Yes\\

    \hline
    \hline
	\end{tabular}
\end{table*}

\begin{table*}
	\caption{
Observations of other three 4FGL sources in MeerKAT searching campaign.}
	\centering
    \renewcommand{\arraystretch}{1.0}
    \setlength\tabcolsep{1.0pt}
    \label{T:undetected_source}
    \renewcommand{\arraystretch}{1.0}
    \setlength\tabcolsep{4.0pt}
	\begin{tabular}{ccc||ccc||ccc}
		\hline
        \hline
        \multicolumn{3}{c}{4FGL~J0523.3$-$2527} & \multicolumn{3}{c}{4FGL~J0940.3$-$7610} & \multicolumn{3}{c}{4FGL~J1120.0$-$2204}\\
        Epoch(MJD) & Band & Number of dishes & Epoch(MJD) & Band & Number of dishes & Epoch(MJD) & Band & Number of dishes \\
        \hline
        59314.55192624 & L & 60 & 59281.71765512 & L & 60 & 59690.92121282 & UHF & 64\\
        59388.20084776 & UHF & 60 & 59383.74660837 & UHF & 56 & 59692.87918223 & L & 60\\
        59692.46091170 & UHF & 60 & 59692.54530037 & UHF & 60 & --- & --- & ---\\
        59698.46137791 & L & 60 & 59698.54582936 & L & 60 & --- & --- & ---\\
    \hline
    \hline
	\end{tabular}
\end{table*}

\bsp	
\label{lastpage}
\end{document}